\documentclass[aps,prd,twocolumn,preprintnumbers,nofootinbib,superscriptaddress,amsmath]{revtex4-2}

\usepackage{graphicx}
\usepackage{amsmath}
\usepackage{amssymb}
\usepackage{amsfonts}
\usepackage[colorlinks]{hyperref}
\usepackage{url}
\usepackage[dvipsnames]{xcolor}
\usepackage{bm}
\usepackage{array}
\usepackage{placeins}
\usepackage{bbold}
\usepackage{algorithm}
\usepackage{algpseudocode}
\usepackage[normalem]{ulem}
\usepackage{xspace}
\usepackage{orcidlink}
\usepackage{subcaption}
\usepackage{multirow}
\usepackage[font=small]{caption}
\usepackage{ragged2e}

\usepackage{tikz}
\usepackage{tikz-3dplot}

\definecolor{dodgerblue}{HTML}{1E90FF}
\definecolor{crimson}{HTML}{DC143C}
\definecolor{auwien}{HTML}{8B64DD}

\hypersetup{
    colorlinks=true,
    linkcolor=auwien,
    filecolor=auwien,
    citecolor = auwien,      
    urlcolor=auwien,
}

\def\d{\mathrm{d}}

\allowdisplaybreaks[4]

\newcommand{\uvec}[1]{\bm{\hat{#1}}}

\newcommand{\ord}[1]{\mathcal{O} \left( #1 \right)}
\newcommand{\D}{\mathcal{D}}
\newcommand{\av}[1]{\left\langle #1 \right\rangle}

\newcommand{\pyEFPEHM}{\texttt{pyEFPEHM}\xspace}

\newcommand{\AEI}{\affiliation{Max Planck Institute for Gravitational Physics (Albert Einstein Institute), D-14476 Potsdam, Germany}}

\newcommand{\IFT}{\affiliation{Instituto de F\'isica Te\'orica UAM/CSIC, Universidad Aut\'onoma de Madrid, Cantoblanco 28049 Madrid, Spain}}

\newcommand{\INTA}{\affiliation{Atmospheric Research and Instrumentation Branch, National Institute for Aerospace Technology (INTA), Madrid, Spain}}

\begin{document}
\makeatother

\title{Horizon absorption in eccentric precessing binary black hole inspirals and its importance for gravitational wave data analysis}

\author{Alberto Álvaro-Díaz \orcidlink{0009-0002-8296-3830}}
\IFT
\INTA

\author{Gonzalo Morras \orcidlink{0000-0002-9977-8546}}
\email{gonzalo.morras@aei.mpg.de}
\AEI
\IFT

\date{\today}

\begin{abstract}

During the evolution of a binary black hole, energy and angular momentum are exchanged between the orbital motion and the individual black holes through horizon absorption, modifying both the binary dynamics and the black hole masses and spins.
This leaves an imprint on the emitted gravitational waves that may be relevant for the accurate modeling of signals observed by current and future detectors, while also offering a probe of the nature of compact objects.
In this work, we derive, for the first time and at leading order in the post-Newtonian expansion, the effect of horizon absorption in binary black hole inspirals with both orbital eccentricity and spin-induced precession, and we incorporate these corrections into the \texttt{pyEFPEHM} waveform model.
We then quantify their impact through analytical estimates of the orbital dephasing, waveform mismatches, and Bayesian parameter-estimation studies.
The effect is largest for systems with large spin components (anti-)aligned with the orbital angular momentum ($|\bm{\chi}_i \cdot \bm{\hat{l}}| \sim 1$), highly unequal mass ratios ($q=m_2/m_1 \ll 1$), and long inspirals spanning a wide frequency range ($\log(f_\mathrm{max}/f_\mathrm{min}) \gg 1$).
For such systems, neglecting horizon absorption biases the recovered binary parameters at moderate signal-to-noise ratios.
In quasi-circular binaries these biases largely absorb the effect, rendering it difficult to detect.
In eccentric binaries, however, the richer signal morphology breaks this degeneracy, making horizon absorption potentially measurable in high signal-to-noise-ratio events.

\end{abstract}

\maketitle

\section{Introduction}
\label{sec:intro}

As the LIGO-Virgo-KAGRA (LVK) Collaboration~\cite{LIGOScientific:2014pky,VIRGO:2014yos,KAGRA:2018plz} continues to observe with improving sensitivity~\cite{KAGRA:2013rdx} and to expand the catalog of detected compact-binary coalescences into the several hundreds~\cite{LIGOScientific:2018mvr,LIGOScientific:2020ibl,KAGRA:2021vkt,LIGOScientific:2025slb,LIGOScientific:2026wfs}, it is becoming increasingly clear that accurate waveform models for generic systems, including orbital eccentricity and spin-induced precession, are required to robustly analyze the observed events and infer their astrophysical formation channels. In particular, evidence for dynamical formation channels has emerged from the observation of very massive black holes with large spins~\cite{LIGOScientific:2020iuh,LIGOScientific:2025rsn}, as well as from lower-mass asymmetric systems with large primary black-hole spins~\cite{LIGOScientific:2025brd}, both suggestive of hierarchical mergers~\cite{Gerosa:2021mno,Alvarez:2024dpd,Li:2025iux}. Similarly, evidence for orbital eccentricity in several high-mass binary black hole mergers~\cite{Gayathri:2020coq,Gamba:2021gap,Romero-Shaw:2022xko,Gupte:2024jfe,Planas:2025jny,Romero-Shaw:2025vbc,Pompili:2026yxq,Lange:2026eqx}, as well as in a neutron star--black hole merger~\cite{Morras:2025xfu,Planas:2025plq,Jan:2025fps,Kacanja:2025kpr,Tiwari:2025fua,Phukon:2025cky}, further supports the presence of dynamically assembled binaries. These observations, together with population studies showing evidence for spin orientations misaligned with the orbital angular momentum~\cite{LIGOScientific:2025pvj} suggest that hierarchical triples~\cite{Stegmann:2025clo,Romero-Shaw:2025otx,Stegmann:2025zkb} undergoing Lidov--Kozai oscillations~\cite{Zeipel:1910,Lidov:1962,Kozai:1962} may contribute to the compact-binary merger population observed by the LVK.

The need for accurate waveform models including eccentricity and precession becomes even stronger for next-generation ground-based detectors such as the Einstein Telescope~\cite{ET:2019dnz,Branchesi:2023mws,ET:2025xjr} and Cosmic Explorer~\cite{Reitze:2019iox,Evans:2021gyd}, as well as space-based observatories such as LISA~\cite{LISA:2017pwj,LISA:2024hlh}. These detectors will probe much lower frequencies, allowing binaries to be observed earlier in their inspiral, where orbital eccentricity is expected to be larger because the binary will have had less time to circularize through gravitational wave (GW) emission. At the same time, the significantly larger signal-to-noise ratios expected in future detections will make physical effects that are subdominant in current LVK observations increasingly important for the unbiased analysis of future detector data. 

One such effect, and the subject of this paper, is horizon absorption~\cite{Price:1986yy,Poisson:1994yf,Tagoshi:1997jy,Alvi:2001mx,Poisson:2004cw}. This effect refers to the absorption (or emission, in the superradiant case~\cite{Press:1972zz,Teukolsky:1974yv,Taracchini:2013wfa,Fujita:2014eta,Gamba:2026fqa}) of gravitational radiation by black hole horizons. In a binary, it leads to an exchange of energy and angular momentum between the orbital motion and the individual black holes, modifying the binary evolution and the black hole masses and spins. While there has been extensive work modeling horizon absorption with different approaches~\cite{Comeau:2009bz,Poisson:2009qj,Poisson:2014gka,Shah:2014tka,Poisson:2018qqd,Munna:2023vds,Warburton:2025ymy,Balivada:2026bdb}, leading to its effect being known to 1.5PN relative order for eccentric-aligned spin binaries~\cite{Chatziioannou:2012gq,Chatziioannou:2016kem,Saketh:2022xjb,Chiaramello:2024unv}, the leading post-Minkowskian order for generic spin scattering orbits~\cite{Jones:2023ugm,Bautista:2024emt,Cipriani:2026myb}, and can be computed numerically at leading order in self-force for generic orbits~\cite{Barack:2018yvs,Pound:2021qin}, analytical results for eccentric-precessing inspirals remain limited.

Horizon absorption is strongly enhanced for spinning black holes, where it enters the binary phase at 2.5PN order, which is significantly less suppressed in the PN expansion than the 4PN correction arising in the non-spinning case. This 2.5PN correction induces a dephasing that accumulates logarithmically with the inspiral duration, making it potentially relevant for sufficiently long signals. Horizon absorption is particularly important in extreme mass-ratio inspirals, where the associated dephasing scales inversely with the binary mass ratio $q = m_2/m_1$~\cite{Bernuzzi:2012ku,Datta:2019epe}. While horizon absorption effects have been observed in numerical relativity simulations of close scattering encounters~\cite{Jaraba:2021ces,Nelson:2019czq,Rodriguez-Monteverde:2024tnt,Rodriguez-Monteverde:2025rfh,Kogan:2025vml,Zhu:2026mhn}, their impact on the waveform is typically small in the comparable-mass, short-duration inspiral-merger-ringdown regime accessible to current simulations. Horizon absorption is incorporated into several state-of-the-art waveform models, including within the effective-one-body framework~\cite{Albanesi:2025txj,Nagar:2011aa,Damour:2012ky}, for extreme-mass-ratio inspirals~\cite{Albertini:2022dmc,Chapman-Bird:2025xtd,Honet:2025gge,Honet:2025lmk,Nishimura:2026nse}, and in phenomenological models~\cite{Mukherjee:2023pge,Mukherjee:2025wxa}. However, it is still neglected in many state-of-the-art waveform models~\cite{Pompili:2023tna,Paul:2024ujx,Planas:2025feq,Hamilton:2025xru,Gamboa:2026jht}, although for those calibrated against numerical simulations the calibration is likely to absorb part of the effect.

Horizon absorption is not only important to model the signals observed by future detectors with sufficient accuracy, but also provides a potential probe of the nature of compact objects, as it is directly tied to the horizon structure of black holes in general relativity and may help distinguish them from exotic compact-object alternatives~\cite{Datta:2019epe,Datta:2020gem,Mukherjee:2022wws,Zi:2023geb,Datta:2024vll}.

In this work, we derive for the first time the effect of horizon absorption in eccentric, precessing binary black hole inspirals, at the leading order in the post-Newtonian (PN) expansion, and incorporate the resulting corrections into the \pyEFPEHM~\cite{Klein:2018ybm,Klein:2021jtd,Morras:2025nlp,Morras:2026fho} waveform model, obtaining an eccentric, precessing inspiral waveform that consistently includes horizon absorption effects. We quantify the impact of horizon absorption through analytical estimates of the orbital dephasing, waveform mismatches, and Bayesian parameter estimation studies. We find that the effect is largest for systems with large spin components aligned or anti-aligned with the orbital angular momentum ($|\bm{\chi}_i \cdot \uvec{l}| \sim 1$), highly unequal mass ratios ($q = m_2/m_1 \ll 1$), and long inspirals spanning a wide frequency range ($\log(f_\mathrm{max}/f_\mathrm{min}) \gg 1$). Neglecting horizon absorption can lead to biased recovery of intrinsic parameters in both quasi-circular and eccentric systems. However, while in the quasi-circular case parameter shifts can largely compensate for the effect, making it difficult to detect directly, in eccentric binaries the richer signal morphology helps break this degeneracy, making horizon absorption potentially measurable in high signal-to-noise ratio events.

The paper is organized as follows. In Sec.~\ref{sec:PNComputations}, we derive the leading-order horizon-absorption corrections for eccentric binaries with generic spin orientations and obtain their orbit-averaged effect on the binary evolution. In Sec.~\ref{sec:Impact}, we estimate the accumulated dephasing, implement the dominant correction in \pyEFPEHM, and quantify its impact through mismatch and parameter-estimation studies. We conclude in Sec.~\ref{sec:conclusion}, summarizing our main results and discussing possible directions for future work.

Unless otherwise specified, we use geometric units ($G=c=1$) and the Euclidean metric for tensor contractions. Vectors are denoted in boldface, with hats indicating unit vectors.

\section{Computation of leading order horizon absorption effects}
\label{sec:PNComputations}

In this section we compute the leading post-Newtonian (PN) effects of horizon absorption in spinning black hole binaries on eccentric orbits, allowing for arbitrary spin orientations.

\subsection{Tidal heating and torquing on generic orbits}
\label{sec:PNComputations:generic}

The evolution of the mass of black hole 1 due to tidal heating from black hole 2 is, at leading PN order, given by~\cite{Chatziioannou:2012gq,Saketh:2022xjb,Chiaramello:2024unv}

\begin{align}
    \frac{\d m_1}{\d t} = m_1^5 \Bigg[& \frac{8}{45} \chi_1 (1 + 3 \chi_1^2) \dot{\mathcal{E}}^{a}{}_b \mathcal{E}_{a c} \hat{S}_1^{b c} \nonumber\\
    & - \frac{2}{3}\chi_1^3 \dot{\mathcal{E}}_{a c} \mathcal{E}_{b d} \hat{s}_1^a \hat{s}_1^b \hat{S}_1^{c d} \Bigg]\, ,
    \label{eq:dm1_dt_raw} 
\end{align}

\noindent where $m_i$ are the component masses, $\chi_i = S_i/m_i^2$ are the dimensionless spin magnitudes, $\hat{s}_i^a$ are unit vectors aligned with the component spins, and $\hat{S}_i^{ab} = \epsilon^{ab}{}_c \hat{s}_i^c$ are the associated spin tensors. The tensor $\mathcal{E}_{ab}$ in Eq.~\eqref{eq:dm1_dt_raw} is the tidal field acting on black hole 1, which for a binary, at leading PN order, is given by~\cite{Poisson:2004cw}

\begin{equation}
    \mathcal{E}_{a b} = \frac{\partial^2}{\partial x^a \partial x^b} \left(- \frac{m_2}{r} \right) = -\frac{m_2}{r^3} \left(3 \frac{x_a x_b}{r^2} - \delta_{a b} \right) \, , 
    \label{eq:Eab}
\end{equation}

\noindent where $x^a$ is the relative position vector and $r = \sqrt{x^a x_a}$. In our leading-order PN computation of tidal torquing and heating, we neglect the contribution from the magnetic tidal field $\mathcal{B}_{ab}$, since its effect enters at relative 1PN order compared to the electric tidal field $\mathcal{E}_{ab}$.

When computing the tidal torquing for binaries with spins misaligned with the orbital angular momentum, both the magnitude and direction of the spin are relevant. From Ref.~\cite{Poisson:2004cw}, the evolution of the spin vector, $S_1^a$, at leading PN order is given by

\begin{equation}
    \frac{\d S_1^a}{\d t} = - \epsilon^a{}_{b c} M_1^{b d} \mathcal{E}_d{}^{c} \, ,
    \label{eq:dS1_dt_raw}    
\end{equation}

\noindent where $\epsilon_{a b c}$ is the Levi-Civita symbol and $M_i^{a b}$ is the black hole mass quadrupole. The latter can be decomposed into an intrinsic contribution, present even for an isolated black hole, and an induced contribution sourced by the external tidal field~\cite{Thorne:1984mz,Poisson:2004cw},

\begin{subequations}
\begin{align}
    M_i^{a b} =& M_{i,\mathrm{intrinsic}}^{a b} + M_{i,\mathrm{induced}}^{a b} \, , \\
    M_{i,\mathrm{intrinsic}}^{a b} =& \frac{1}{3} m_i^3 \chi_i^2 \left(\delta^{a b} - 3 \hat{s}_i^a \hat{s}_i^b\right) \, , \\ 
     M_{i,\mathrm{induced}}^{a b} =& \frac{2}{45} m_i^5 \chi_i \Bigg[\lambda(\chi_i) \left(\delta^{a b} - 3 \hat{s}_i^a \hat{s}_i^b\right) \mathcal{E}_{c d} \hat{s}_i^{c} \hat{s}_i^d \nonumber\\
     & + 8 (1 + 3 \chi_i^2) \mathcal{E}^{c (a}\epsilon^{b)}{}_{c d} \hat{s}_i^d  \nonumber\\
     & + 30 \chi_i^2 \hat{s}_i^{(a}\epsilon^{b)}{}_{c d} \hat{s}_i^c \mathcal{E}^{d}{}_e \hat{s}_i^e  \Bigg] \, ,
\end{align}
\end{subequations}

\noindent where $v^{(a} v^{b)} = (v^a v^b + v^b v^a)/2$ denotes symmetrization. The contribution from the intrinsic quadrupole is already accounted for in the spin precession equations~\cite{Thorne:1984mz}, and will therefore be neglected in what follows. Furthermore, the coefficient $\lambda(\chi_i)$, related to the tidal deformability of the black hole and left undetermined in Ref.~\cite{Poisson:2004cw}, was later shown to vanish in General Relativity~\cite{Chia:2020yla,Charalambous:2021mea}. Both of these contributions are proportional to $\delta^{a b} - 3 \hat{s}_i^a \hat{s}_i^b$, and substituting them into Eq.~\eqref{eq:dS1_dt_raw} yields a purely conservative spin evolution of the form $\dot{S}_1^a = \epsilon^a{}_{b c} \Omega^b S_1^c$, which does not affect the secular phasing of the binary. Since contributions from the mass evolution of the companion ($\propto \dot{m}_2$) enter at higher PN orders, they are neglected.

Taking these simplifications into account, and using in Eqs.~\eqref{eq:dm1_dt_raw} and~\eqref{eq:dS1_dt_raw} the explicit form of the leading-order tidal field in Eq.~\eqref{eq:Eab}, we obtain the leading-order tidal heating and torquing for generic orbits,

\begin{subequations}
\label{eq:dm1dS1_gen}
\begin{align}
    \frac{\d m_1}{\d t} =& - K_1 \frac{\epsilon_{a b c} \hat{s}_1^a x^b \dot{x}^c}{r^2} \, , \label{eq:dm1dS1:dm1} \\
    \frac{\d S_1^a}{\d t} =& - K_1 \left[\hat{s}_1^a - \frac{\hat{s}_1^b x_b}{r} \frac{x^a}{r} \right] \, , \\
    K_1 =& \frac{8}{5} \frac{m_1^5 m_2^2}{r^6} \chi_1 \left\{1 + 3 \left[1 - \frac{5}{4} \left(\frac{\hat{s}_1^d x_d}{r}\right)^2 \right] \chi_1^2 \right\} \, ,
\end{align}
\end{subequations}

\noindent where the results for the second black hole can be obtained by exchanging labels $1 \leftrightarrow 2$. Eq.~\eqref{eq:dm1dS1_gen} agrees with the result derived in Ref.~\cite{Goldberger:2020fot} using world-line effective field theory.

\subsection{Tidal heating and torquing on eccentric orbits}
\label{sec:PNComputations:eccentric}

To compute the tidal heating and torquing at leading order, we model the motion using Newtonian (0PN) orbits. In doing so, we assume that the orbital period is much shorter than the precession and radiation-reaction timescales. This is justified at the leading PN order considered here, since precession and radiation reaction are 1.5PN and 2.5PN corrections to the orbit, respectively, and therefore enter at the same relative order in the effect we compute.

For bound Newtonian orbits, the motion is given by the Keplerian parametrization~\cite{Damour:1985ecc,Colwell:1993book}

\begin{subequations}
\label{eq:NewtonianOrbit}
\begin{align}
    r(u) & = a (1 - e \cos{u}) \, ,\label{eq:NewtonianOrbit:r} \\
    \phi(u) & = v(u) \equiv 2 \arctan\left[\left(\frac{1+e}{1-e}\right)^{1/2} \tan{\frac{u}{2}}\right] \, , \label{eq:NewtonianOrbit:v} \\
    \ell(u) & \equiv n (t - t_0) = u - e \sin{u} \, , \label{eq:NewtonianOrbit:l}
\end{align}
\end{subequations}

\noindent where $a$ is the semi-major axis, $e$ the eccentricity (with $0 \leq e <1 $), $n = 2 \pi/P$ the mean motion (with $P$ the orbital period), and $t_0$ is a constant of integration; the auxiliary variables $u$, $v$ and $\ell$ are the eccentric, true and mean anomalies, respectively.

We parametrize the orbit using the eccentricity $e$ and the PN parameter

\begin{equation}
    y = \frac{(M \omega)^{1/3}}{\sqrt{1 - e^2}} \, , 
    \label{eq:y_def}
\end{equation}

\noindent where $M = m_1+m_2$ is the total mass, $\omega$ is the orbital frequency and each power of $y$ corresponds to a relative 0.5PN order in the PN expansion. These quantities are related to the binding energy $E < 0$ and angular momentum $L$ by

\begin{subequations}
\label{eq:ye_of_EL}
\begin{align}
    y =& \frac{M^2 \nu}{L} \, , \\
    e^2 =& 1 + \frac{2 E L^2}{M^5 \nu^3} \, ,
\end{align}
\end{subequations}

\noindent where $\nu = m_1 m_2/M^2$ is the symmetric mass ratio. The orbital elements in Eq.~\eqref{eq:NewtonianOrbit} can then be written as~\cite{Boetzel:2017zza}

\begin{subequations}
\label{eq:an_of_ye}
\begin{align}
    a =& \frac{M}{(1-e^2) y^2} \, , \\
    n =& \frac{(1-e^2)^{3/2}y^3}{M} \, ,
\end{align}
\end{subequations}

\begin{figure}[t!]
    \centering  
    \includegraphics[width=0.49\textwidth]{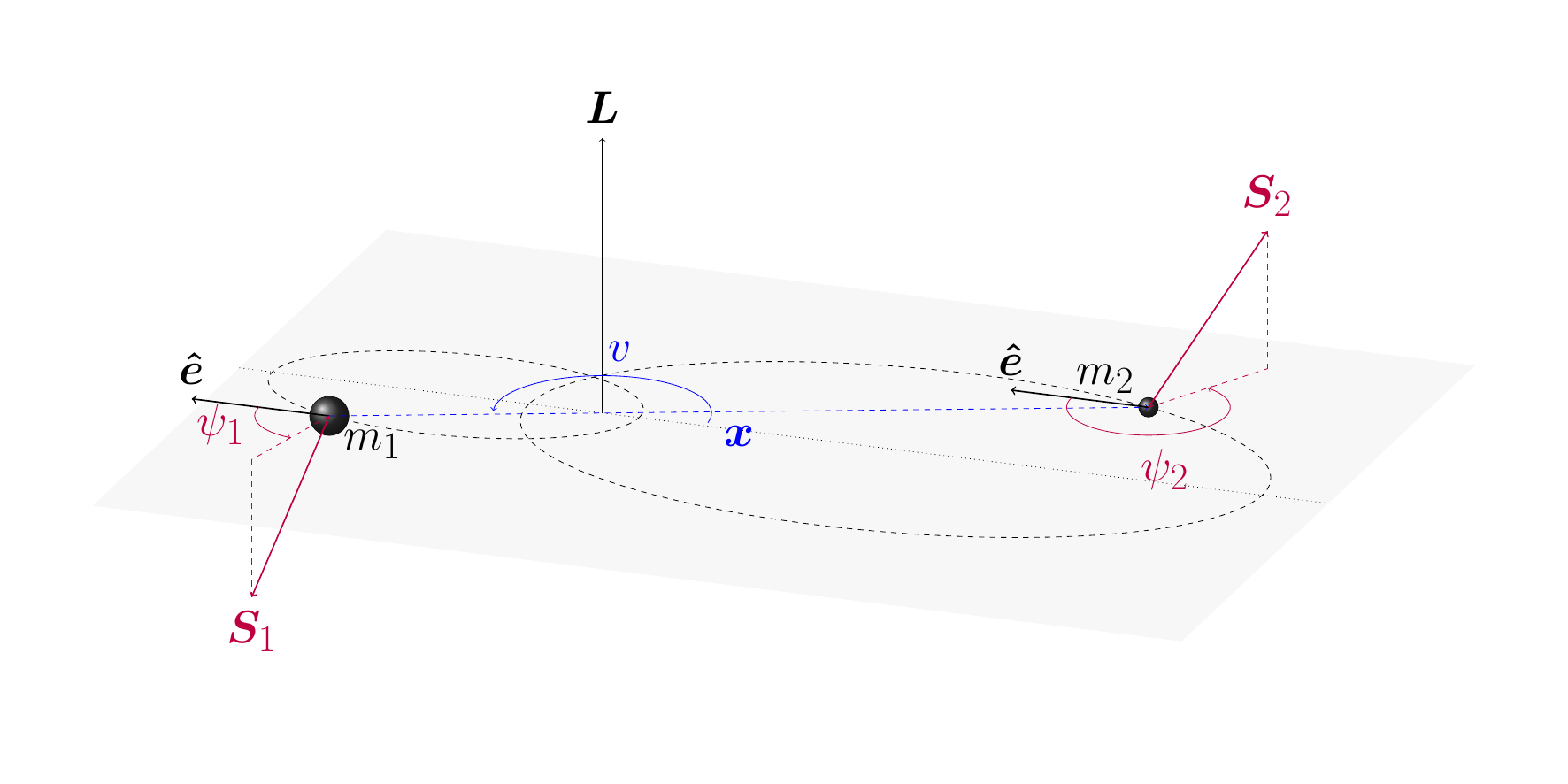}
    \caption{\justifying Schematic representation of the binary configuration. The orbital plane is perpendicular to the orbital angular momentum vector. The angles $\psi_i$ are defined as the angles between the projection of the spin vectors onto the orbital plane and the Laplace-Runge-Lenz vector, defined in Eq.~\eqref{eq:LRL_vec}.}
    \label{fig:binary_fig}
\end{figure}

To carry out the computation, we choose coordinates such that the relative separation vector is given by

\begin{align}
    \bm{x} =& r\left(\cos{v},\,\sin{v}, \, 0\right) \nonumber\\
    =& a \left(\cos{(u)} - e, \, \sqrt{1-e^2}\sin{(u)}, \, 0 \right) \,
    \label{eq:x_kep}
\end{align}

\noindent and the spin directions are parametrized as

\begin{equation}
    \uvec{s}_i = \left(\sin{\theta_i}\cos{\psi_i},\, \sin{\theta_i} \sin{\psi_i},\, \cos{\theta_i}  \right) \, ,
    \label{eq:si_kep}
\end{equation}

\noindent where, comparing with Eq.~\eqref{eq:x_kep}, $\theta_i$ is the angle between the spin vector and the orbital angular momentum, and $\psi_i$ is the angle between the projection of the spin onto the orbital plane and the periastron direction. A schematic representation of the binary configuration and the relevant angles is shown in Fig.~\ref{fig:binary_fig}.

Substituting Eqs.~\eqref{eq:x_kep} and~\eqref{eq:si_kep} into Eq.~\eqref{eq:dm1dS1_gen}, we can obtain the instantaneous tidal heating and torquing. However, for waveform modeling, the small periodic variations in the spin and mass within each orbit are subdominant compared to the secular changes accumulated over many cycles. To extract this secular contribution, we perform an orbital average defined by

\begin{align}
    \av{f}_\mathrm{orb} &= \frac{1}{P} \int_{t}^{t+P} \d t' \; f(t') = \frac{1}{2 \pi} \int_{-\pi}^\pi \d \ell \; f(\ell) \nonumber \\
    & = \frac{1}{2 \pi} \int_{-\pi}^\pi \d u \; (1 - e \cos{u}) f(u) \, ,
    \label{eq:orb_avg_def}    
\end{align}

\noindent where $f$ represents an arbitrary periodic function with period $P$ and Eq.~\eqref{eq:NewtonianOrbit:l} has been used. Performing the orbital average of the instantaneous tidal heating and torquing, we obtain 

\begin{widetext}
\begin{subequations}
\label{eq:dm1dS1_OrbAvg}
\begin{align}
    \av{\D m_i}_\mathrm{orb} =& -M \nu^2 \mu_i^3 y^{15} (\uvec{l} \cdot \bm{\chi}_i) C_{\D m_i} \, , \\
    \av{\D \bm{S}_i}_\mathrm{orb} =& -M^2 \nu^2 \mu_i^3 y^{12} \Big\{ C_{\D S_i,\chi} \, \bm{\chi}_i - C_{\D S_i,\chi_\perp} \, \big[\bm{\chi}_i - (\uvec{l} \cdot \bm{\chi}_i) \uvec{l}\big]  -  C_{\D S_i,e} \, (\uvec{e} \cdot \bm{\chi}_i) \uvec{e} \Big\} \, , \label{eq:dm1dS1_OrbAvg:dS1} \\
    C_{\D m_i} =& \left(\frac{8}{5}+12 e^2+9 e^4+\frac{e^6}{2}\right) \left[1 +  \frac{9}{8} \chi_i^2 + \frac{15}{8} (\uvec{l} \cdot \bm{\chi}_i)^2 \right] - e^2 \left(\frac{45}{4}+\frac{45 e^2}{4}+\frac{45 e^4}{64}\right) |\uvec{l} \times \bm{\chi}_i|^2 \cos{2 \psi_i} \, , \\
    C_{\D S_i,\chi} =& \left(\frac{8}{5}+\frac{24 e^2}{5}+\frac{3 e^4}{5}\right) \left[1 +  \frac{9}{8} \chi_i^2 + \frac{15}{8} (\uvec{l} \cdot \bm{\chi}_i)^2 \right] - e^2 \left(\frac{9}{2}+\frac{3 e^2}{4}\right) |\uvec{l} \times \bm{\chi}_i|^2 \cos{2 \psi_i} \, , \\
    C_{\D S_i,\chi_\perp} =& \left( \frac{4}{5}+\frac{6 e^2}{5}+\frac{e^4}{10} \right) (1 + 3 \chi_i^2) - |\uvec{l} \times \bm{\chi}_i|^2 \left[ \frac{9}{4}+\frac{9 e^2}{2}+\frac{27 e^4}{64}+e^2 \left(\frac{9}{4}+\frac{9 e^2}{32}\right) \cos{2 \psi_i} \right] \, , \\
    C_{\D S_i,e} =& e^2 \left\{\left(\frac{12}{5}+\frac{2 e^2}{5}\right) \left[1 +  \frac{9}{8} \chi_i^2 + \frac{15}{8} (\uvec{l} \cdot \bm{\chi}_i)^2 \right] -\frac{3 e^2}{16} |\uvec{l} \times \bm{\chi}_i|^2 \cos{2 \psi_i} \right\} \, , 
\end{align}
\end{subequations}
\end{widetext}

\noindent where $\mu_i =m_i/M$ are the dimensionless mass parameters and $\bm{\chi}_i = \chi_i \uvec{s}_i = \bm{S}_i/m_i^2$ is the dimensionless spin vector, we introduce the derivative operator

\begin{equation}
    \D = \frac{M}{\left(1-e^2 \right)^{3/2}} \frac{d}{dt} \, ,
    \label{eq:D_def}
\end{equation}

\noindent $\uvec{l}$ is the unit vector along the orbital angular momentum, 

\begin{equation}
    \uvec{l} = \frac{\bm{x} \times \dot{\bm{x}}}{|\bm{x} \times \dot{\bm{x}}|} = (0, \, 0, \, 1) \, ,
    \label{eq:l_def}
\end{equation}

\noindent and the eccentricity vector $\uvec{e} = \bm{e}/e$ is the unit vector along the periastron line, with $\bm{e}$ the Laplace-Runge-Lenz vector, defined as

\begin{equation}
    \bm{e} = \frac{\dot{\bm{x}} \times (\bm{x} \times \dot{\bm{x}})}{M} - \frac{\bm{x}}{r} = (e,\, 0,\, 0) \, .
    \label{eq:LRL_vec}
\end{equation}

Note that the eccentricity vector $\uvec{e}$ is ill-defined in the quasi-circular limit ($e=0$). However, Eq.~\eqref{eq:dm1dS1_OrbAvg} shows that the component of $\av{\D \bm{S}_i}_\mathrm{orb}$ along $\uvec{e}$ vanishes as $e \to 0$, consistent with the absence of a preferred periastron direction.

To our knowledge, this is the first derivation of tidal heating and torquing for precessing binaries. In the spin-aligned limit, Eq.~\eqref{eq:dm1dS1_OrbAvg} reduces to previously known results in the literature, both for quasi-circular~\cite{Alvi:2001mx,Poisson:2004cw,Taylor:2008xy,Chatziioannou:2012gq,Saketh:2022xjb} and eccentric orbits~\cite{Datta:2023wsn}.

\subsection{Evolution of the orbits}
\label{sec:PNComputations:orbs}

The exchange of energy and angular momentum between the black holes and the orbit modifies not only the masses and spins of the black holes, but also the orbital dynamics, which determines the GW phase evolution, the primary observable in the signal. The variation of any orbital quantity can be computed using the chain rule,

\begin{align}
    \D f = \frac{\partial f}{\partial E} \D E + \frac{\partial f}{\partial \bm{L}} \cdot \D \bm{L} + \sum_{i=1}^2 \left( \frac{\partial f}{\partial m_i} \D m_i + \frac{\partial f}{\partial \bm{S}_i} \D \bm{S}_i \right) .
    \label{eq:D_orbparam}
\end{align}

To obtain the horizon-absorption contributions to $\D E$ and $\D \bm{L}$, we use the conservation of energy and angular momentum, which implies

\begin{subequations}
\label{eq:DEDL_of_DmDS}
\begin{align}
    (\D E)_H \equiv & -\mathcal{F}_H = - \D m_1 - \D m_2 \, , \\
    (\D \bm{L})_H \equiv & -\bm{\mathcal{G}}_H = - \D \bm{S}_1 - \D \bm{S}_2 \, , 
\end{align}
\end{subequations}

\noindent where $\mathcal{F}_H$ and $\bm{\mathcal{G}}_H$ denote the energy and angular momentum fluxes. Combining Eqs.~\eqref{eq:ye_of_EL}, \eqref{eq:dm1dS1_OrbAvg}, \eqref{eq:D_orbparam}, and \eqref{eq:DEDL_of_DmDS}, we obtain the leading-order horizon-flux contributions to the evolution of the PN parameter and eccentricity,

\begin{widetext}
\begin{subequations}
\label{eq:DyDe_OrbAvg}
\begin{align}
    (\D y)_H =& -\nu \mu_1^3 y^{14} (\uvec{l} \cdot \bm{\chi}_1) \left\{ \left(\frac{8}{5}+\frac{24 e^2}{5}+\frac{3 e^4}{5}\right) \left[1 +  \frac{9}{8} \chi_1^2 + \frac{15}{8} (\uvec{l} \cdot \bm{\chi}_1)^2 \right] - e^2 \left(\frac{9}{2}+\frac{3 e^2}{4}\right) |\uvec{l} \times \bm{\chi}_1|^2 \cos{2 \psi_1} \right\} \nonumber \\
    & + \left( 1 \leftrightarrow 2 \right) , \\
    (\D e^2)_H =& \nu \mu_1^3 y^{13} (\uvec{l} \cdot \bm{\chi}_1) e^2  \left\{ \left(\frac{88}{5}+\frac{132 e^2}{5}+\frac{11 e^4}{5}\right) \left[1 +  \frac{9}{8} \chi_1^2 + \frac{15}{8} (\uvec{l} \cdot \bm{\chi}_1)^2 \right] - \left(\frac{27}{2}+30 e^2+\frac{93 e^4}{32}\right) |\uvec{l} \times \bm{\chi}_1|^2 \cos{2 \psi_1} \right\} \nonumber\\
    &+ \left( 1 \leftrightarrow 2 \right) , 
\end{align}
\end{subequations}
\end{widetext}

\noindent where we have omitted the orbit-average brackets for notational simplicity. Note that, in deriving Eq.~\eqref{eq:DyDe_OrbAvg}, the evolution of the masses and spins (i.e., the terms proportional to $\partial f/\partial m_i$ and $\partial f/\partial \bm{S}_i$ in Eq.~\eqref{eq:D_orbparam}) does not contribute at leading order, entering only at relative 1PN order. In the spin-aligned limit, the expressions in Eq.~\eqref{eq:DyDe_OrbAvg} reduce to known results in the literature~\cite{Datta:2023wsn,Henry:2023tka}.

In Eq.~\eqref{eq:DyDe_OrbAvg} we observe that, for unequal masses, the effect of horizon absorption is strongly dominated by the absorption of the primary black hole, with the contribution of the secondary suppressed by a factor $\propto q^3$ for equal dimensionless spins.

\section{Impact of horizon absorption on waveform modeling}
\label{sec:Impact}

In this section we assess the impact of horizon absorption on waveform modeling. We begin with an analytical estimate of the size of the effect, showing that its dominant observable signature is a 2.5PN modification of the GW phasing. We then incorporate this effect into \pyEFPEHM~\cite{Morras:2026fho}, obtaining an eccentric precessing inspiral waveform model with horizon absorption, which we use to study its impact through mismatch and parameter-estimation studies.

\subsection{Analytical estimate of horizon absorption effects}
\label{sec:Impact:Estimate}

To quantify the effects of horizon absorption, we use that the evolution of the PN parameter and the eccentricity can schematically be written as~\cite{Klein:2018ybm}

\begin{subequations}
\label{eq:DyDe2_gen}
\begin{align}
    \D y =& \nu y^9 \sum_{n \geq 0} a_n(\nu, y, e^2, \uvec{l}, \bm{\chi}_1, \bm{\chi}_2) y^n \, , \label{eq:DyDe2_gen:Dy} \\
    \D e^2 =& -\nu y^8 \sum_{n \geq 0} b_n(\nu, y, e^2, \uvec{l}, \bm{\chi}_1, \bm{\chi}_2) y^n \, ,
\end{align}
\end{subequations}

\noindent where $a_n$ and $b_n$ are dimensionless coefficients of order $\ord{\nu^0 y^0}$, whose explicit dependence on $y$ is limited to terms proportional to $\log{y}$ starting at 3PN~\cite{Klein:2018ybm}. In particular, the leading (0PN) order coefficients are given by~\cite{Peters:1964zz}

\begin{subequations}
\label{eq:a0b0}
\begin{align}
    a_0 =& \frac{32}{5} + \frac{28}{5} e^2  \, , \\
    b_0 =&  \frac{608}{15} e^2 + \frac{242}{15} e^4 \, .
\end{align}
\end{subequations}

The leading-order evolution of the masses and spins as a function of the PN parameter $y$ can be obtained by dividing Eqs.~\eqref{eq:dm1dS1_OrbAvg} by Eq.~\eqref{eq:DyDe2_gen:Dy}. The total change during the inspiral then follows by integrating between an initial and final value, $y_0$ and $y_f$. For generic eccentricity, these integrals do not admit simple closed-form analytical expressions. However, we can show that the result is dominated by the quasi-circular contribution. To do so, we note that the integrands can be expanded in eccentricity as

\begin{equation}
    \frac{\d f}{\d y} = y^{n-1} \sum_{m \geq 0} f_m e^{2m} \, .
    \label{eq:dfdy_ye2_expanded}
\end{equation}

To integrate Eq.~\eqref{eq:dfdy_ye2_expanded} we need an expression for $e^2(y)$, found by solving Eq.~\eqref{eq:DyDe2_gen}. At 0PN this can be done analytically, obtaining

\begin{equation}
    e^2 = e_0^2 \left(\frac{1 + \frac{121}{304} e^2}{1 + \frac{121}{304} e_0^2} \right)^{-145/121} \left(\frac{y}{y_0}\right)^{-19/3}   \, ,
    \label{eq:e2_0PN_decay}
\end{equation}

\noindent where $e_0$ and $y_0$ are the eccentricity and PN parameter at some initial reference time. From Eq.~\eqref{eq:e2_0PN_decay} we can construct an upper bound on $e^2$ during the inspiral,

\begin{align}
    e^2 \leq \overline{e}_0^2  \left(\frac{y}{y_0}\right)^{-19/3}  \, ,
    \label{eq:e2_decay_UpperBound}
\end{align}

\noindent with 

\begin{equation}
    \overline{e}_0^2 = e_0^2 \left(1 + \frac{121}{304} e_0^2 \right)^{145/121} \, .
    \label{eq:e02_bar}
\end{equation}

Although here we work at leading order, we note that Ref.~\cite{Morras:2026fho} showed that Eq.~\eqref{eq:e2_decay_UpperBound} remains valid up to 3PN order when neglecting the residual eccentricity induced by precession~\cite{Klein:2010ti}. The bound of Eq.~\eqref{eq:e2_decay_UpperBound} can be used to derive an upper bound to the contribution of each term in Eq.~\eqref{eq:dfdy_ye2_expanded}, noting that

\begin{align}
    I_{n m} & \equiv \int_{y_0}^{y_f} e^{2m} y^{n-1}\d y \leq \overline{e}_0^{2 m} \int_{y_0}^{y_f} \left(\frac{y}{y_0}\right)^{-\frac{19}{3}m} y^{n-1}\d y \nonumber \\
    & = \frac{\overline{e}_0^{2 m} y_0^n}{\frac{19}{3}m - n} \left[1 - \left(\!\frac{y_0}{y_f}\!\right)^{\!\!\frac{19}{3}m - n}  \right] ,
    \label{eq:Inm_def}
\end{align}

\noindent and using that the final PN parameter is much larger than the initial one ($y_f \gg y_0$), we have that

\begin{align}
    I_{n m} \lesssim 
    \begin{cases}
        \frac{\overline{e}_0^{2 m}}{n - \frac{19}{3} m} y_0^{\frac{19}{3} m} y_f^{n - \frac{19}{3} m} & m < \frac{3}{19}n  \\[0.5em]
        \overline{e}_0^{2 m} y_0^{\frac{19}{3} m} \log{\frac{y_f}{y_0}} & m = \frac{3}{19} n \\[0.5em]
        \frac{\overline{e}_0^{2 m}}{\frac{19}{3} m - n}  y_0^{n + 1} & m > \frac{3}{19} n
    \end{cases}  \, .
    \label{eq:Inm_y0_ll_yf}
\end{align}

Therefore, we find that, for $n>0$, all eccentric ($m>0$) contributions are suppressed by powers of the small parameter $y_0$, and the integral is therefore dominated by the quasi-circular ($m=0$) terms. The intuitive explanation is that eccentricity is radiated away faster than high-PN contributions can accumulate. For the evolution of the mass $\d m_i/\d y$ and spin $\d \bm{S}_i/\d y$, we have that $n = 7$ and $4$, respectively, and therefore only the quasi-circular contributions are expected to be significant.

To integrate Eqs.~\eqref{eq:dm1dS1_OrbAvg} and obtain the secular changes in the masses and spins, we must account for spin precession. The precession equations for an eccentric system at next-to-leading order (NLO), which we denote as 2PN to match the PN counting used for the orbital phase evolution, are given by~\cite{Barker:1979jmv,Racine:2008qv,Klein:2018ybm,Klein:2021jtd}:

\begin{subequations}
\label{eq:raw_prec_eqs}
\begin{align}
\D \uvec{L} &= - y^6 \left( \bm{\Omega}_1 + \bm{\Omega}_2 \right) \, , \label{eq:raw_prec_eqs:L} \\
\D \bm{s}_1 &= \mu_2 y^5 \bm{\Omega}_1 \, ,  \label{eq:raw_prec_eqs:s1} \\
\D \bm{s}_2 &= \mu_1 y^5 \bm{\Omega}_2 \, ,  \label{eq:raw_prec_eqs:s2}
\end{align}
\end{subequations}

\noindent where

\begin{subequations}
\label{eq:raw_prec_eqs_defs}
\begin{align}
\bm{s}_i &= \mu_i \bm{\chi}_i =  \frac{\bm{S}_i}{M^2 \mu_i} \, , \\
\bm{\Omega}_i &=  \left[ \frac{1}{2}\mu_i + \frac{3}{2} \left(1 - y \chi_\mathrm{eff} \right) \right] \uvec{l} \times \bm{s}_i + \frac{1}{2} y \bm{s}_j \times \bm{s}_i \, , \\
\chi_\mathrm{eff}  &= \uvec{l}\cdot (\bm{s}_1 + \bm{s}_2) \, .
\end{align}
\end{subequations}

Comparing Eq.~\eqref{eq:raw_prec_eqs} with Eqs.~\eqref{eq:dm1dS1_OrbAvg}, we find that the contribution to $\D \bm{s}_i$ from horizon absorption enters at relative 5.5PN order (i.e., N$^7$LO). This is well beyond the highest order currently available, which reaches 4PN (i.e., N$^4$LO) for quasi-circular binaries~\cite{Bohe:2012mr,Sturani:2015STA,Akcay:2020qrj,Khalil:2023kep}.

The precession equations in Eq.~\eqref{eq:raw_prec_eqs} imply that the spins and the orbital angular momentum precess about the direction of the total angular momentum $\uvec{\jmath} = \bm{J}/J$, where

\begin{equation}
    \bm{J} = \bm{L} + \bm{S}_1 + \bm{S}_2 \, ,
    \label{eq:J_LO_def}
\end{equation}

\noindent and the direction of $\uvec{\jmath}$ is approximately conserved over the inspiral.

The characteristic timescale of the precessional motion is $T_\mathrm{p} \sim \ord{y^{-5}}$, which is much shorter than the radiation-reaction timescale $T_\mathrm{RR} \sim \ord{y^{-8}}$. This separation of timescales allows us to average the rapid precessional motion over the slower radiation-reaction evolution.

Since the spins undergo rapid precessional motion in the orbital plane, the angle $\psi_i$ between the in-plane spin and the periastron line increases monotonically, with $\D \psi_i \sim \ord{y^5}$, even when including the effects of periastron advance~\cite{Klein:2021jtd}, which implies that

\begin{equation}
    \av{f(\uvec{l},\bm{\chi}_i) \cos{2 \psi_i}} = 0.
\end{equation}

\noindent where $\av{\cdot}$ denotes an average over the precessional timescale and $f(\uvec{l},\bm{\chi}_i)$ is an arbitrary scalar function of $\uvec{l}$ and $\bm{\chi}_i$.

Similarly, when computing the spin evolution due to horizon absorption by integrating Eq.~\eqref{eq:dm1dS1_OrbAvg:dS1}, the rapid precession causes the components perpendicular to $\uvec{\jmath}$ to average out. As a result, the secular spin evolution is aligned with $\uvec{\jmath}$, and we obtain

\begin{equation}
    \av{\D \bm{S}_i} = \av{\uvec{\jmath} \cdot \D \bm{S}_i} \uvec{\jmath} \, .
    \label{eq:dS1_precavg}
\end{equation}

Taking into account this precession average, together with the arguments around Eq.~\eqref{eq:Inm_y0_ll_yf}, the dominant contribution to the tidal heating and torquing between an initial and final PN parameter $y_0$ and $y$ is given by the quasi-circular, precession-averaged contribution. At leading PN order, we obtain

\begin{subequations}
\label{eq:Delta_mi_Si}
\begin{align}
    \Delta m_i =& - \frac{1}{28} M \nu \av{\kappa_{H,i}} \left[y^7 - y_0^7 + \ord{e_0^2 y_0^{19/3} y^{2/3}}\right] \nonumber\\
    & + \ord{y^9} , \\
    \Delta \bm{S}_i  =& - \frac{1}{16} M^2 \nu \av{\kappa_{H,i}} \left[y^4 - y_0^4 + \ord{e_0^2 y_0^4}\right] \uvec{\jmath} \nonumber \\
    & + \ord{y^6} , \\
    \kappa_{H,i} \equiv & \, \mu_i^3 (\uvec{l}\cdot\bm{\chi}_i) \left[1 +  \frac{9}{8} \chi_i^2 + \frac{15}{8} (\uvec{l} \cdot \bm{\chi}_i)^2 \right] \, ,
\end{align}
\end{subequations}

\noindent where, in performing the integration, we have treated the masses and precession-averaged quantities as constant, since their evolution with $y$ due to horizon absorption and radiation reaction is PN suppressed.

From Eq.~\eqref{eq:Delta_mi_Si}, we see that the changes in the masses and spins due to tidal heating and torquing enter at relative 3.5PN and 2PN orders, respectively. Since the masses and spins enter the orbital evolution equations~\eqref{eq:DyDe2_gen} at 0PN and 1.5PN order, both effects induce corrections to the phasing at relative 3.5PN order. This is one PN order higher than the leading contribution from horizon absorption to the orbital evolution, given in Eq.~\eqref{eq:DyDe_OrbAvg}.

Although the variations of the masses and spins formally enter at the same order as the 1PN corrections to Eq.~\eqref{eq:DyDe_OrbAvg}, their impact on the phasing is expected to be subleading. This is because these effects carry an additional suppression by the symmetric mass ratio, and therefore do not grow in the extreme mass-ratio limit ($\nu \to 0$). 

Moreover, the small numerical prefactors in Eq.~\eqref{eq:Delta_mi_Si} further suppress their contribution, implying that the effect of mass and spin evolution due to horizon absorption remains negligible compared to the direct flux contributions, even away from the extreme mass-ratio regime.

Therefore, the dominant impact of horizon absorption on the waveform is a dephasing induced by the direct horizon flux contributions in Eq.~\eqref{eq:DyDe_OrbAvg}. To estimate this effect, we use that, from Eq.~\eqref{eq:y_def}, the mean orbital phase $\lambda$ evolves as

\begin{equation}
    \D\lambda = y^3 \, .
    \label{eq:Dlambda}
\end{equation}

The dephasing $\Delta\lambda = \lambda - \lambda_0$ induced by an additional contribution $\delta(\D y)$ to a reference evolution $(\D y)_0$ can then be written as

\begin{equation}
    \frac{\d \Delta \lambda}{\d y} = \frac{\D\lambda}{(\D y)_0 + \delta(\D y)} - \frac{\D\lambda}{(\D y)_0} \, ,
    \label{eq:DDeltalambda}
\end{equation}

\noindent where $\delta(\D y)$ may also include contributions induced by modifications to the eccentricity evolution $e^2(y)$ arising from corrections to $\D e^2$. Assuming $\delta(\D y) \ll (\D y)_0$, we expand to first order and obtain

\begin{equation}
    \frac{\d \Delta \lambda}{\d y} = - \frac{y^3}{(\D y)_0^2}\delta(\D y) \left[1 + \ord{\frac{\delta(\D y)}{(\D y)_0}} \right] \, .
    \label{eq:DDeltalambda_expanded}
\end{equation}

Using for $\delta(\D y)$ the horizon absorption contribution in Eq.~\eqref{eq:DyDe_OrbAvg}, and for $(\D y)_0$ the leading-order evolution from Eqs.~\eqref{eq:DyDe2_gen} and~\eqref{eq:a0b0}, together with the precession averaging and the arguments around Eq.~\eqref{eq:Inm_y0_ll_yf}, we find

\begin{equation}
    \Delta \lambda_H = \frac{5}{256} \frac{\langle \tilde{\kappa}_H \rangle}{\nu} \left[ \log{\frac{y}{y_0}} + \ord{e_0^2 y^0} \right] + \ord{y^2} ,
    \label{eq:Deltalambda_LO}
\end{equation}

\noindent where we have defined

\begin{align}
    \tilde{\kappa}_H \equiv & 2 \left(\kappa_{H,1} + \kappa_{H,2} \right) \nonumber\\
    =& \left[1 - 2 \nu + \frac{9}{8} \left(s_1^2 + s_2^2\right) \right] \chi_\mathrm{eff} + \left[\delta\mu +\frac{9}{8}\left(s_1^2 - s_2^2\right)\right]\delta\chi \nonumber\\
    & + \frac{45}{16} \chi_\mathrm{eff} \delta\chi^2 + \frac{15}{16}\chi_\mathrm{eff}^3 \, ,
    \label{eq:kappa_H_def}    
\end{align}

\noindent with

\begin{subequations}
    \begin{align}
        \delta\mu &= \mu_1 - \mu_2 \, , \\
        \chi_\mathrm{eff} &= \uvec{l} \cdot (\bm{s}_1 + \bm{s}_2) \, , \\ 
        \delta\chi &= \uvec{l} \cdot (\bm{s}_1 - \bm{s}_2) \, .
    \end{align}
\end{subequations}

From Eq.~\eqref{eq:Deltalambda_LO}, we observe that the leading-order horizon absorption effect can accumulate an arbitrarily large dephasing for sufficiently long inspirals, growing logarithmically with $y/y_0$. Furthermore, since $\Delta \lambda_H \propto 1/\nu$, the dephasing is enhanced in the extreme mass-ratio limit ($\nu \to 0$).

We also observe that eccentric corrections are not entirely negligible. While they do not grow large with $y$, they accumulate to a constant contribution of order $\ord{e_0^2/\nu}$, which can become relevant in the extreme mass-ratio limit, and should therefore be included in waveform models.

Finally, we note that the overall prefactor in Eq.~\eqref{eq:Deltalambda_LO} is numerically small. As a result, accumulating an $\ord{1}$ dephasing from horizon absorption requires very long inspirals ($y \gg y_0$) or highly asymmetric mass ratios ($\nu \ll 1$). In particular, comparing the contributions from Eq.~\eqref{eq:DyDe_OrbAvg} to the 2.5PN spin-orbit terms~\cite{Morras:2025nlp}, the horizon-absorption terms are smaller by a factor of $\sim 100$.

\subsection{Adding horizon absorption to \pyEFPEHM}
\label{sec:Impact:pyEFPE}

In \pyEFPEHM, we include the dominant effect of horizon absorption, namely the modification to the PN evolution equations induced by the direct horizon flux contributions in Eq.~\eqref{eq:DyDe_OrbAvg}. Higher-order effects, such as the evolution of the component masses and spins, are neglected, as they enter at higher PN order and have a subdominant impact on the waveform, as discussed in Sec.~\ref{sec:Impact:Estimate}, and would significantly complicate the implementation. For computational efficiency, \pyEFPEHM employs precession-averaged equations of motion. The precession-averaged horizon absorption contributions to the PN evolution coefficients in Eq.~\eqref{eq:DyDe2_gen} are given by

\begin{subequations}
\label{eq:a5Hb5H_EFPE}
\begin{align}
    a_5^H &= -\left(\frac{4}{5} + \frac{12}{5} e^2 + \frac{3}{10} e^4 \right) \av{\tilde{\kappa}_H} \, , \\
    b_5^H &= -e^2 \left(\frac{44}{5} + \frac{66}{5} e^2 + \frac{11}{10} e^4\right) \av{\tilde{\kappa}_H} \, ,
\end{align}
\end{subequations}

\noindent where $\tilde{\kappa}_H$ is defined in Eq.~\eqref{eq:kappa_H_def}, and the precession averages of the quantities entering it are already computed within \pyEFPEHM~\cite{Morras:2025nlp,Morras:2026fho}.

\begin{figure}[t!]
\centering  
\includegraphics[width=0.49\textwidth]{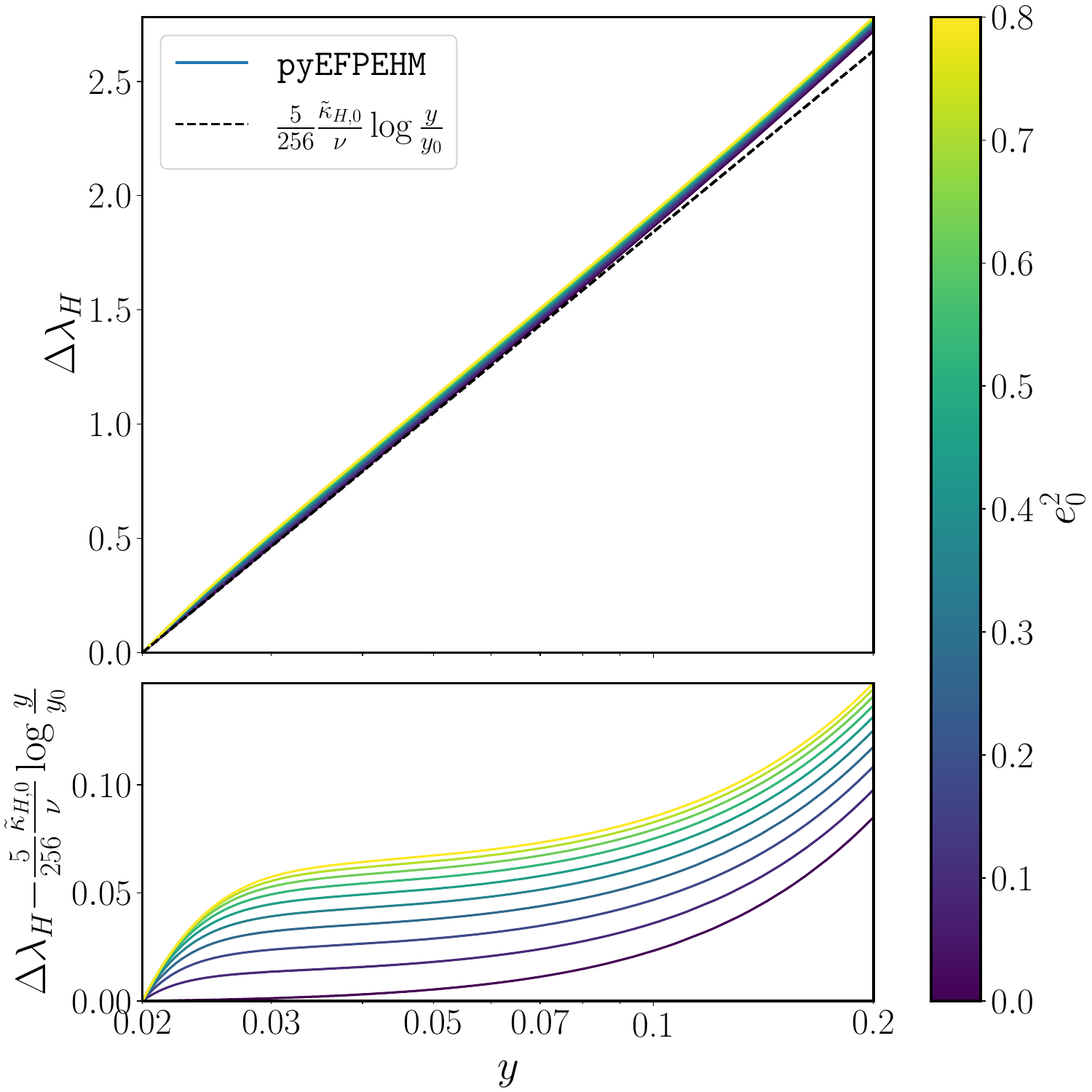}
\caption{\justifying Comparison of the orbital phase difference $\Delta\lambda_H$ between \pyEFPEHM waveforms with and without horizon absorption and the leading-order quasi-circular prediction of Eq.~\eqref{eq:Deltalambda_LO}, as a function of the PN parameter $y$. Different colors correspond to different values of the initial squared eccentricity $e_0^2$. We consider a fiducial system with mass ratio $q = m_2/m_1 = 0.1$, and initial spins $\bm{\chi}_{1,0} = [-0.2, 0.3, 0.9]$ and $\bm{\chi}_{2,0} = [-0.4, -0.4, 0.8]$, chosen so that horizon absorption is large while the binary still precesses. In the top panel we show $\Delta\lambda_H$ from \pyEFPEHM together with the analytical prediction of Eq.~\eqref{eq:Deltalambda_LO}, while in the bottom panel we show their difference.
}
\label{fig:DeltaLambda_EFPE_exp_of_y}
\end{figure}

After implementing the leading-order horizon absorption effects in \pyEFPEHM, we compare the dephasing predicted by the model with the analytical estimate of Eq.~\eqref{eq:Deltalambda_LO} to assess the validity of some of the approximations introduced in Sec.~\ref{sec:Impact:Estimate}. This is shown in Fig.~\ref{fig:DeltaLambda_EFPE_exp_of_y}, where we plot the orbital phase difference $\Delta\lambda_H$ between \pyEFPEHM waveforms with and without horizon absorption, compared with Eq.~\eqref{eq:Deltalambda_LO}, as a function of the PN parameter $y$ and the initial eccentricity $e_0$.

We consider a fiducial system with mass ratio $q = m_2/m_1 = 0.1$, and initial spins $\bm{\chi}_{1,0} = [-0.2, 0.3, 0.9]$ and $\bm{\chi}_{2,0} = [-0.4, -0.4, 0.8]$, defined in a frame where $\uvec{l} = [0,0,1]$, as in \pyEFPEHM. In Eq.~\eqref{eq:Deltalambda_LO}, we evaluate $\tilde{\kappa}_H$ using the initial spin configuration ($\tilde{\kappa}_{H,0} \approx 4.84$), rather than its precession average.

This configuration is not typical of the observed astrophysical population~\cite{LIGOScientific:2025pvj}, but is deliberately chosen so that horizon absorption is large while the binary still precesses appreciably, combining a highly asymmetric mass ratio with near-extremal spins at moderate tilts.

We find good agreement between the analytical estimate and \pyEFPEHM. In particular, consistently with the discussion in Sec.~\ref{sec:Impact:Estimate}, the dephasing in \pyEFPEHM is dominated by the quasi-circular contribution. As the eccentricity increases, the deviation from Eq.~\eqref{eq:Deltalambda_LO} grows proportionally to $e_0^2$ at low eccentricities, and rapidly saturates to a constant as a function of $y$.

Finally, even in the quasi-circular case ($e_0 = 0$), we observe a residual difference that grows as $\sim y^2$. This behavior is consistent with both the neglect of 1PN corrections in $(\D y)_0$ when deriving Eq.~\eqref{eq:Deltalambda_LO} from Eq.~\eqref{eq:DDeltalambda_expanded}, and the approximation of the time-dependent $\av{\tilde{\kappa}_H}$ by its initial value $\tilde{\kappa}_{H,0}$.

\subsection{Measuring the impact of horizon absorption}
\label{sec:Impact:Measure}

A rough estimate for horizon absorption to have a measurable impact is that it induces a significant orbital dephasing, $\Delta\lambda_H \sim \ord{1}$. As shown in Fig.~\ref{fig:DeltaLambda_EFPE_exp_of_y}, the orbital dephasing is well described by Eq.~\eqref{eq:Deltalambda_LO}, which can be written in terms of the initial and final orbital frequency and eccentricity as

\begin{align}
    \Delta \lambda_H \approx \frac{5}{768} \frac{\langle \tilde{\kappa}_H \rangle}{\nu} \log\left\{\frac{f_f}{f_0} \left(\frac{1 - e_0^2}{1 - e_f^2} \right)^{3/2} \right\} \, .
    \label{eq:Deltalambda_LO_physical}
\end{align}

\begin{figure}[t!]
\centering  
\includegraphics[width=0.49\textwidth]{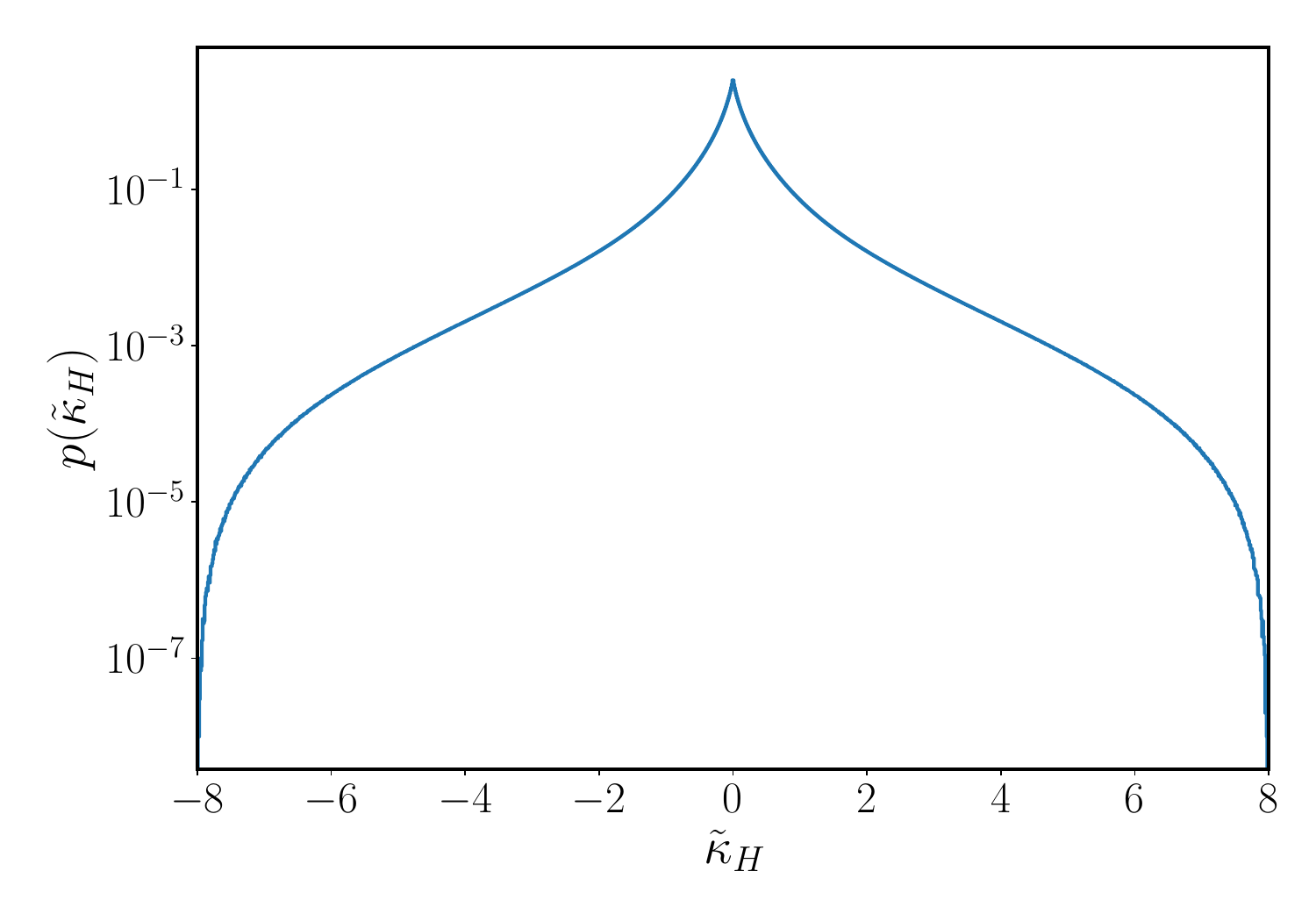}
\caption{\justifying Probability density function for $\tilde{\kappa}_H$. The distribution is obtained by evaluating Eq.~\eqref{eq:kappa_H_def} for $10^{10}$ samples with parameters drawn from uniform distributions in the mass ratio $q \in [0, 1]$ and spin magnitudes $\chi_i \in [0, 1]$, with isotropic spin orientations.
}
\label{fig:p_kH_q_0_1_chi_0_1_N_1e+10}
\end{figure}

The dephasing is proportional to $\tilde{\kappa}_H$, defined in Eq.~\eqref{eq:kappa_H_def}. While $|\tilde{\kappa}_H|$ can reach values as large as $|\tilde{\kappa}_H| = 8$ in the limit $q \to 0$ with maximally (anti-)aligned spins, it is typically smaller. To quantify this, Fig.~\ref{fig:p_kH_q_0_1_chi_0_1_N_1e+10} shows the probability density function of $\tilde{\kappa}_H$ assuming uniform distributions in $q \in [0,1]$ and $\chi_i \in [0,1]$, with isotropic spin orientations. The distribution is symmetric and sharply peaked at $\tilde{\kappa}_H = 0$. It has standard deviation $\mathrm{STD}[\tilde{\kappa}_H] \approx 0.724$, indicating that $\tilde{\kappa}_H$ is typically an $\ord{1}$ quantity. Large values $|\tilde{\kappa}_H| \sim 8$ are extremely rare, as they require both highly asymmetric mass ratios and maximal spins with finely tuned (anti-)aligned spin configurations.

Assuming a quasi-circular system ($e_0 = e_f = 0$), equal masses ($\nu = 1/4$), and an observation spanning two orders of magnitude in frequency ($f_f \sim 10^2 f_0$), as typical for broadband GW detectors, we obtain the following order-of-magnitude estimate on the orbital dephasing

\begin{align}
    \Delta \lambda_H \approx 0.12 \tilde{\kappa}_H \frac{1/4}{\nu} \frac{\log\left\{\frac{f_f}{f_0} \left(\frac{1 - e_0^2}{1 - e_f^2} \right)^{3/2} \right\}}{\log{10^2}} \, .
    \label{eq:Deltalambda_estimate}
\end{align}

Under these conservative assumptions, we see that the orbital dephasing is close to $\ord{1}$ even when $\tilde{\kappa}_H \sim \ord{1}$. Relaxing these assumptions, the effect can be further enhanced for sufficiently unequal mass ratios, due to the $1/\nu$ scaling, and for longer inspirals, due to the $\log(f_f/f_0)$ accumulation.

In Eq.~\eqref{eq:Deltalambda_estimate}, we observe that the orbital dephasing between two orbital frequencies $f_0$ and $f_f$ nominally decreases as the initial eccentricity increases. Nonetheless, as the system becomes more eccentric, higher-order harmonics are excited, such that the orbital frequencies accessible at a fixed GW frequency scale as $\propto (1 - e^2)^{-3/2}$~\cite{Morras:2025nlp}. As a result, the observable orbital dephasing remains approximately constant. Moreover, the presence of multiple harmonics helps break parameter degeneracies and improves the measurability of the dephasing, implying that smaller values of $\Delta\lambda$ may be detectable~\cite{Morras:2025nlp}.

These analytical considerations suggest that horizon absorption may be detectable in favorable configurations, particularly for systems with long observable inspirals, unequal mass ratios, and relatively large values of $|\tilde{\kappa}_H|$. However, an orbital dephasing of $\Delta\lambda_H \sim \ord{1}$ is only a rough criterion for detectability, and full waveform models for eccentric precessing inspirals exhibit a rich structure where different physical effects can interplay with horizon absorption.

To assess the actual impact of horizon absorption on GW observations, we now turn to quantitative measures based on waveform mismatches and parameter-estimation studies.

\subsubsection{Mismatch studies}
\label{sec:Impact:Measure:Mismatch}

While the orbital dephasing provides useful intuition about the distinguishability between waveforms, a more quantitative measure of waveform difference is the mismatch. The relevant quantity to compare is the detector strain, which, in the long-wavelength approximation~\cite{Schutz:1987xok} can be written as

\begin{equation}
    h = F_+(\alpha,\delta,\psi) h_+ + F_\times(\alpha,\delta,\psi) h_\times \, ,
    \label{eq:h_detector}
\end{equation}

\noindent where $F_+$ and $F_\times$ are the antenna pattern functions, which depend on the detector geometry and on the sky location $(\alpha,\delta)$ and polarization angle $\psi$. For convenience, this detector response can be recast in terms of an effective polarization angle $\kappa(\alpha, \delta, \psi)$ as~\cite{Ramos-Buades:2023ehm}

\begin{equation}
    h = A(\alpha,\delta,\psi) \big[ \cos{(2\kappa)} h_+ + \sin{(2\kappa)} h_\times \big]\, .
    \label{eq:h_from_pols}
\end{equation}

Given two strains $h_1$ and $h_2$, their mismatch $\mathcal{MM}$~\cite{Sathyaprakash:1991mt,Finn:1992xs} is defined as
\begin{equation}
    \mathcal{MM}(h_1, h_2) = 1 - \mathcal{M}(h_1, h_2) = 1 - \frac{\langle h_1 | h_2 \rangle}{\sqrt{\langle h_1 | h_1 \rangle \langle h_2 | h_2 \rangle}} \, ,
    \label{eq:SimpleMisMatchDef}
\end{equation}

\noindent where $\mathcal{M}(h_1, h_2)$ is the match, and the noise-weighted inner product $\langle \cdot | \cdot \rangle$ is defined as

\begin{equation}
    \langle a | b \rangle = 4 \mathrm{Re}\left\{ \int_{f_\mathrm{min}}^{f_\mathrm{max}} \frac{\tilde{a}^{*}(f) \tilde{b}(f)}{S_n(f)} \, \d f \right\} \, ,
    \label{eq:InnerProdDef}
\end{equation}

\noindent where $f_\mathrm{min}$ and $f_\mathrm{max}$ are the minimum and maximum frequencies analyzed, and $S_n(f)$ is the one-sided noise power spectral density (PSD) of the detector.

A commonly used criterion for distinguishability is that, for two signals to be observationally distinguishable, their mismatch must satisfy~\cite{Lindblom:2008cm}

\begin{equation}
    \mathcal{MM} \gtrsim \frac{N_p}{2 \rho^2} \, ,
    \label{eq:distinguishable_MM}
\end{equation}

\noindent where $\rho$ is the signal-to-noise ratio (SNR) and $N_p$ is the effective number of parameters describing the signal.

While the mismatch provides a necessary (though not sufficient) condition for distinguishing two waveforms, parameter biases depend on the detailed structure of waveform errors and on whether they can be absorbed by shifts in the inferred parameters~\cite{Thompson:2025hhc}. In practice, part of the waveform differences can be absorbed by shifts in parameters that are not of interest, such as the orbital phase $\phi_0$, the coalescence time $t_0$, the polarization angle $\kappa$, the mean anomaly $\ell_0$, or rigid rotations of the in-plane spins by an angle $\phi_S$. To obtain a measure that is more directly related to potential biases in the parameters of interest, we minimize the mismatch over these ``nuisance'' parameters. We thus define the minimized mismatch as

\begin{equation}
    \overline{\mathcal{MM}}(h_1, h_2) = \min_{\substack{\phi_0, t_0, \kappa,\\ \ell_0, \phi_S}} \mathcal{MM}(h_1, h_2)  \, .
    \label{eq:MinMisMatchDef}
\end{equation}

Following Ref.~\cite{Harry:2016ijz}, the minimization over the polarization angle $\kappa$ is performed analytically, while the minimization over the reference time $t_0$ is efficiently carried out using the fast Fourier transform (FFT) followed by a local refinement using Brent's method~\cite{Brent:1971}. The minimization over the reference phase $\phi_0$, the mean anomaly $\ell_0$ and the in-plane spin rotation angle $\phi_S$ is performed numerically using a Bayesian optimization algorithm~\cite{Jones1998}, which we make publicly available in~\cite{pybop_repo}.

\begin{figure}[t!]
\centering  
\includegraphics[width=0.49\textwidth]{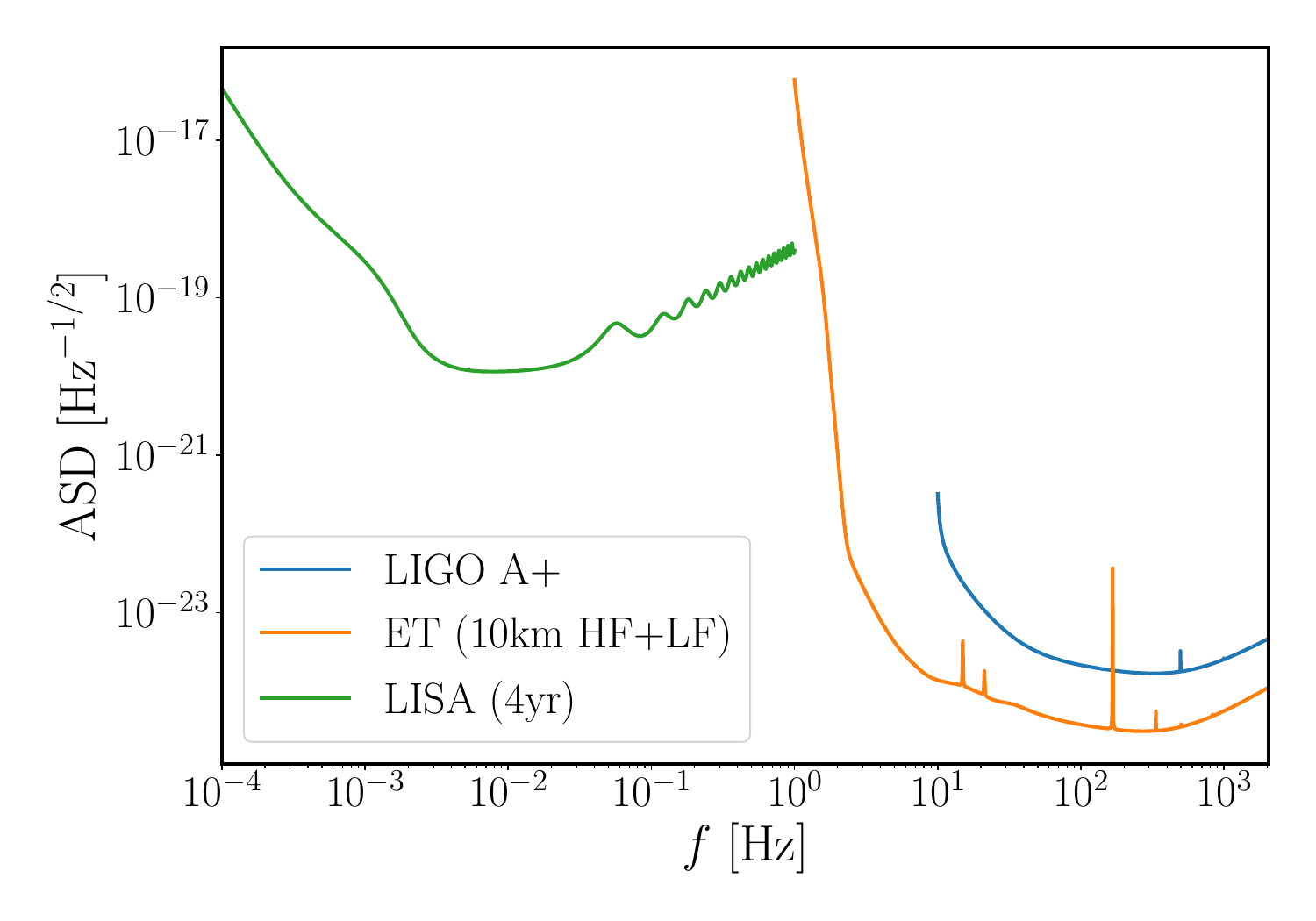}
\caption{\justifying Noise ASDs used in the mismatch computations for Advanced LIGO A+, the Einstein Telescope (ET), and LISA, plotted over the corresponding frequency ranges employed in the analysis.
}
\label{fig:ASDs_pyEFPE_comparison_q_e}
\end{figure}

Here, we compute mismatches for representative ground- and space-based detectors, namely Advanced LIGO, the Einstein Telescope (ET), and LISA. In each case, the noise-weighted inner product of Eq.~\eqref{eq:InnerProdDef} is evaluated using the corresponding PSD, integrating over the frequency range where the detector is sensitive. The amplitude spectral densities ($\mathrm{ASD}=\sqrt{\mathrm{PSD}}$) used in our analysis, plotted over the frequency ranges used in the mismatch computation, are shown in Fig.~\ref{fig:ASDs_pyEFPE_comparison_q_e}.

For LIGO, we use the projected ``Advanced LIGO A+'' PSD~\cite{KAGRA:2013rdx,ObservingScenariosPSDs}, and perform the integration over the frequency range $f \in [10\,\mathrm{Hz}, 2048\,\mathrm{Hz}]$.

For ET, we use the PSD corresponding to a 10 km triangular configuration with both high- and low-frequency interferometers (``ET 10km HF+LF'' in Ref.~\cite{ET_PSDs_CoBA}), and perform the integration over the frequency range $f \in [1\,\mathrm{Hz}, 2048\,\mathrm{Hz}]$.

For LISA, we use the analytic noise model described in Ref.~\cite{Robson:2018ifk}, assuming a mission duration of $4\,\mathrm{yr}$. This model effectively includes instrumental noise, the detector response, and the Galactic confusion background. The integration is performed over the frequency range $f \in [10^{-4}\,\mathrm{Hz}, 1\,\mathrm{Hz}]$.

In all cases, the waveforms are generated up to the innermost stable circular orbit~\cite{Morras:2025nlp}. For LIGO and ET, waveforms start when the $l=m=n=2$ main GW mode of the binary equals the minimum analysis frequency. For LISA, however, we sometimes choose a higher initial frequency to ensure that the signal duration does not exceed the assumed mission lifetime of $4\,\mathrm{yr}$.

For each interferometer, the chirp mass range explored is chosen such that the signals contain a significant inspiral phase where \pyEFPEHM is applicable. In particular we study $\mathcal{M}_c \in [1, 10] M_\odot$ in Advanced LIGO A+, $\mathcal{M}_c \in [1, 100] M_\odot$ in ET and $\mathcal{M}_c \in [10, 10^5] M_\odot$ in LISA. Therefore, some of the signals studied are extremely long, making the mismatch computation computationally expensive. To address this, we minimize the mismatch over $\phi_0$, $\ell_0$, and $\phi_S$ using a Bayesian optimization algorithm~\cite{Jones1998} with support for periodic parameters, which reliably finds the minimum with $\mathcal{O}(50)$ evaluations.

\begin{figure}[t!]
    \centering
    \includegraphics[width=0.49\textwidth]{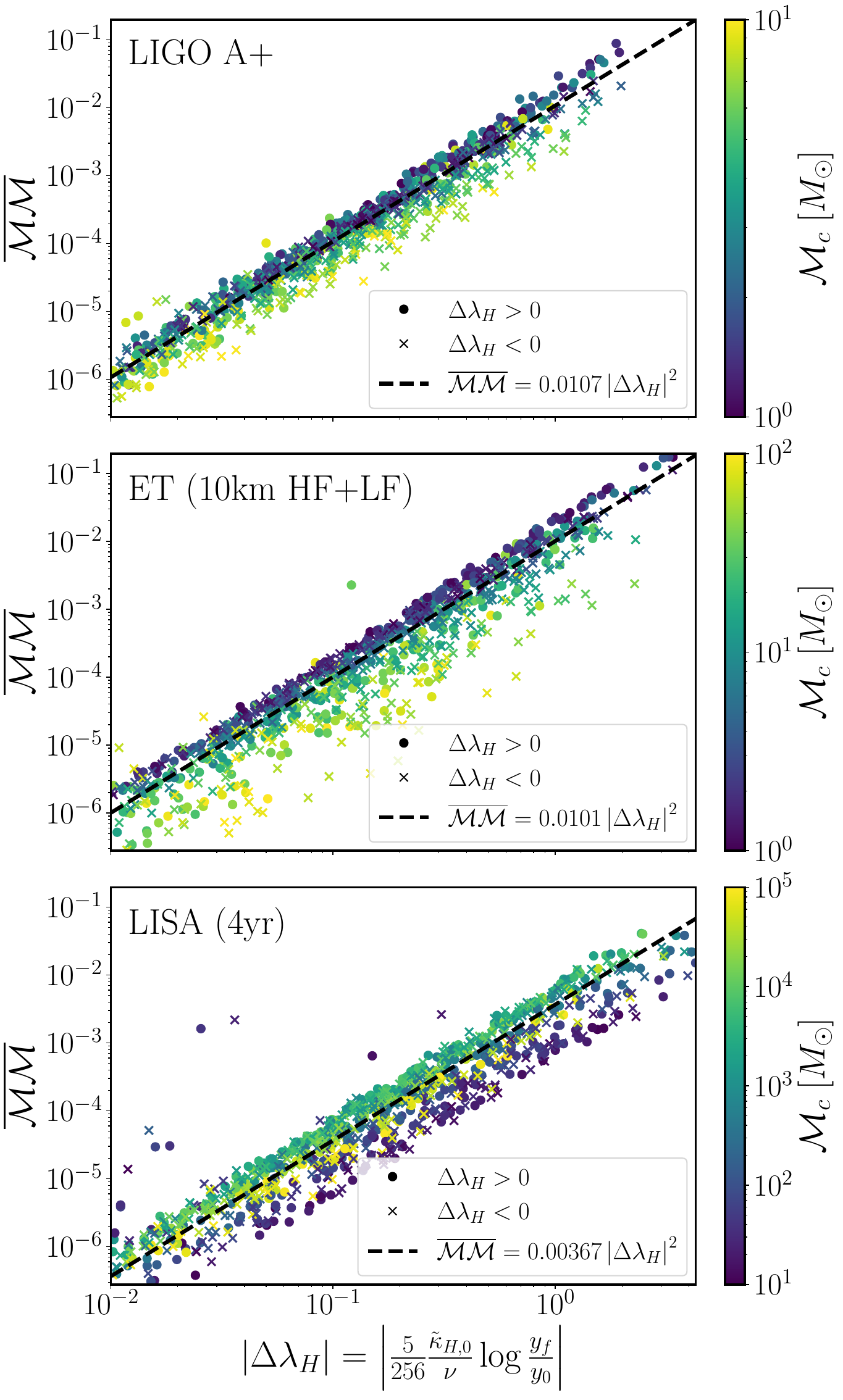}
    \caption{\justifying Scatter plots of the minimized mismatches $\overline{\mathcal{MM}}$ as a function of the absolute value of the approximate horizon absorption dephasing $|\Delta \lambda_H|$ of Eq.~\eqref{eq:Deltalambda_LO}. The top, middle, and bottom panels correspond to Advanced LIGO A+, ET, and LISA, respectively. In all panels, the color of the points indicates the chirp mass in solar masses ($M_\odot$).
    }
    \label{fig:MM_absdeltalambda_3ifos_N_1024}
\end{figure}

We start by exploring the relation between the minimized mismatch and the approximate horizon absorption dephasing derived in Eq.~\eqref{eq:Deltalambda_LO}. To this end, we compute mismatches between \pyEFPEHM waveforms with and without horizon absorption for random samples uniformly distributed in the logarithm of the chirp mass, the inverse mass ratio $1/q \in [1, 20]$, the component spin magnitudes $\chi_i \in [0, 1]$, and the initial eccentricity $e_0 \in [0, 0.6]$, with isotropic binary orientations and spin directions. In Fig.~\ref{fig:MM_absdeltalambda_3ifos_N_1024} we show, for the three detectors studied, the mismatch of the samples against their approximate dephasing computed using Eq.~\eqref{eq:Deltalambda_LO}. We observe that in all cases the mismatch is strongly correlated with the approximate dephasing, with

\begin{equation}
    \overline{\mathcal{MM}} \propto |\Delta \lambda_H|^2 \, ,
    \label{eq:MM_dphase_relation}
\end{equation}

\noindent with detector-dependent proportionality constants given in Fig.~\ref{fig:MM_absdeltalambda_3ifos_N_1024}. This is consistent with the expectation that the mismatch is proportional to the square of the waveform dephasing~\cite{Lindblom:2008cm} and that, as we saw in Sec.~\ref{sec:Impact:Estimate}, Eq.~\eqref{eq:Deltalambda_LO} for $\Delta \lambda_H$ is a good estimate of the dephasing. Therefore, the dephasing of Eq.~\eqref{eq:Deltalambda_LO} can serve as a simple proxy for whether or not horizon absorption is detectable, even in the case of eccentric precessing binaries.

In Fig.~\ref{fig:MM_absdeltalambda_3ifos_N_1024} we observe that the correlation between the mismatch and the dephasing is not perfect. For the same dephasing the mismatch can vary by more than an order of magnitude based on the other binary parameters.

\begin{figure*}
    \centering
    \includegraphics[width=\textwidth]{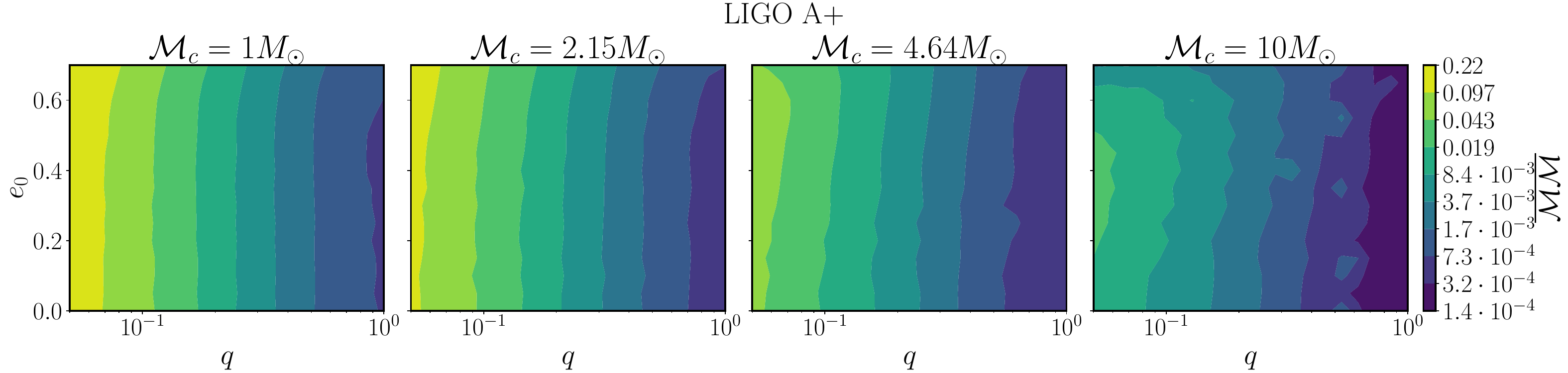}
    \includegraphics[width=\textwidth]{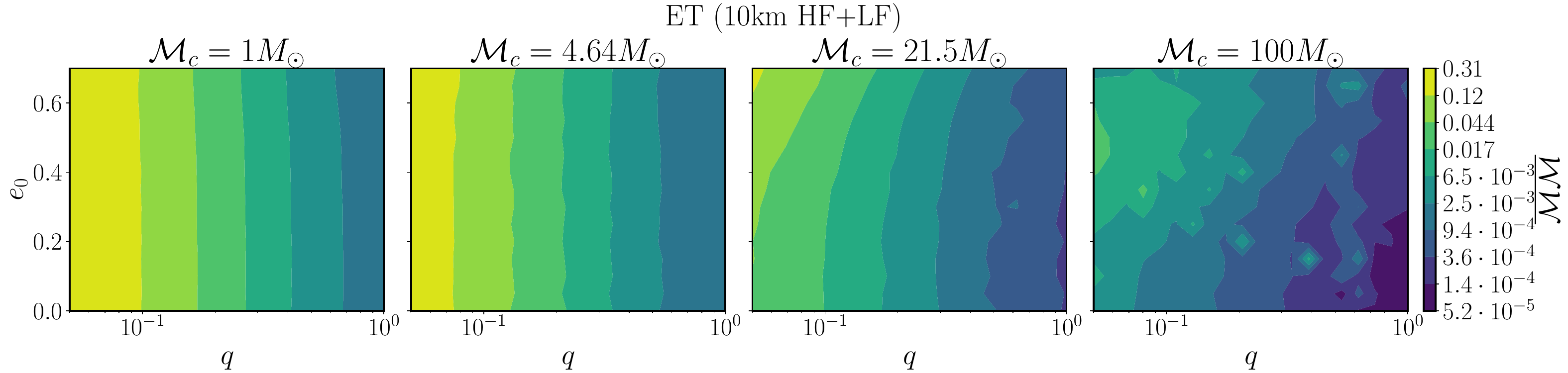}
    \includegraphics[width=\textwidth]{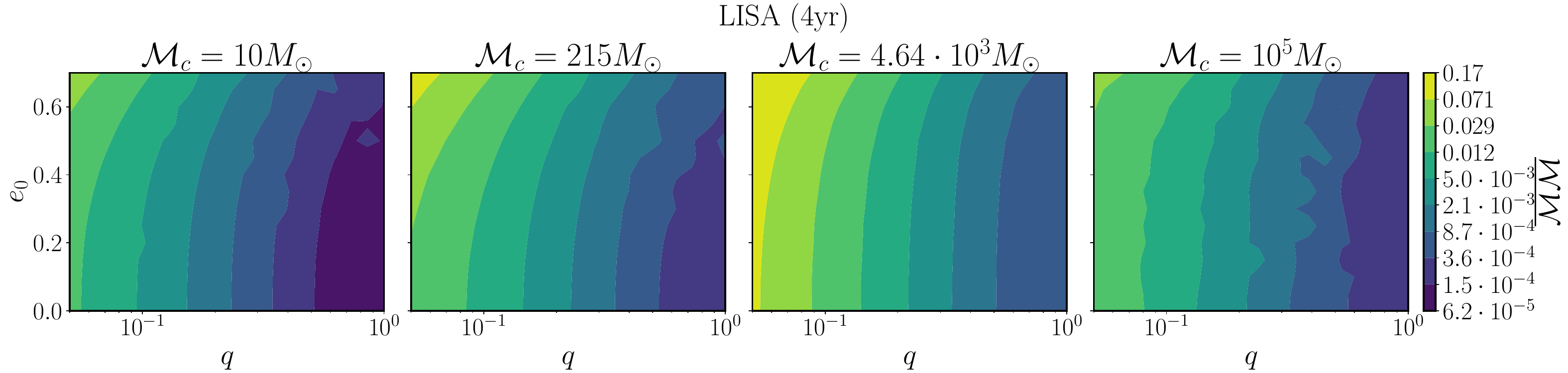}
    \caption{\justifying Minimized mismatch $\overline{\mathcal{MM}}$ as a function of mass ratio $q = m_2/m_1 \in [0.05, 1]$ and initial eccentricity $e_0 \in [0, 0.7]$ for different chirp masses and detectors. Each panel corresponds to a different choice of chirp mass (indicated in the panel title). The top, middle, and bottom sets of panels correspond to Advanced LIGO A+, ET, and LISA, respectively. The spins are fixed to the fiducial configuration $\bm{\chi}_{1,0} = [-0.2, 0.3, 0.9]$ and $\bm{\chi}_{2,0} = [-0.4, -0.4, 0.8]$, for which horizon absorption is large.
    }
    \label{fig:mismatches_pyEFPE_of_q_e_compare_HorizonAbsorption}
\end{figure*}

To better understand the dependence of the mismatch on the physical binary parameters, we study the mismatch as a function of the chirp mass, mass ratio, and initial eccentricity, while fixing the remaining parameters to fiducial values. In particular, we adopt the same spin configuration as in Fig.~\ref{fig:DeltaLambda_EFPE_exp_of_y}, namely $\bm{\chi}_{1,0} = [-0.2, 0.3, 0.9]$ and $\bm{\chi}_{2,0} = [-0.4, -0.4, 0.8]$. We further fix the initial mean anomaly $\ell_0 = 4.3$, inclination $\iota_0 = \pi/3$, reference phase $\phi_0 = 1.2$, and effective polarization $\kappa = \pi/2$.

The results for the three interferometers are shown in Fig.~\ref{fig:mismatches_pyEFPE_of_q_e_compare_HorizonAbsorption}, where we plot the minimized mismatch $\overline{\mathcal{MM}}$ as a function of the mass ratio $q$ and the initial eccentricity $e_0$ for several choices of the chirp mass. We observe that the mismatch depends most strongly on the mass ratio, increasing for more unequal systems, consistent with the $1/\nu$ scaling of the dephasing induced by horizon absorption (see Eq.~\eqref{eq:Deltalambda_LO_physical}).

For LIGO and ET, the mismatch increases for smaller chirp masses. This can be understood from two effects: lower-mass systems merge at higher frequencies, enhancing the $\log(f_f/f_0)$ factor in Eq.~\eqref{eq:Deltalambda_LO_physical}, and they accumulate a larger number of cycles, making it more difficult for the nuisance parameters to absorb the induced dephasing. For LISA the behavior is more nuanced. For high masses ($\mathcal{M}_c \gtrsim 10^3$), the same trends and reasoning apply. However, for lower masses, where the inspiral proceeds more slowly, the signal duration is limited by the mission lifetime ($4\,\mathrm{yr}$). Therefore, the systems start being observed from a higher starting frequency $f_0$, reducing the $\log(f_f/f_0)$ factor in Eq.~\eqref{eq:Deltalambda_LO_physical} and thus decreasing the mismatch.

The dependence on eccentricity is more subtle. On the one hand, increasing eccentricity reduces the dephasing predicted by Eq.~\eqref{eq:Deltalambda_LO_physical}. On the other hand, higher eccentricities excite additional harmonics, which enhance the sensitivity to phase differences by effectively providing multiple phase tracers. Intuitively, as the signal becomes more burst-like, a small shift in the time of periastron passage can lead to noticeable mismatches. At the same time, increasing eccentricity shortens the signal, reducing the number of cycles and allowing nuisance parameters to absorb part of the dephasing.

As a result of these competing effects, we find that eccentricity generally increases the mismatch for long signals (low chirp mass and moderate eccentricity), while decreasing it when the signals become too short (high chirp mass and very large eccentricity).

We observe that, across detectors, the mismatch ranges from $\sim 0.3$ for unequal-mass systems observed over a broad frequency range to $\sim 5 \times 10^{-5}$ for near-equal-mass systems observed over a narrower frequency interval. Using the distinguishability criterion of Eq.~\eqref{eq:distinguishable_MM}, this corresponds to distinguishable SNRs ranging from $\rho \sim \mathcal{O}(5)$ for the largest mismatches to $\rho \gtrsim \mathcal{O}(400)$ for the smallest ones, assuming a typical number of effective parameters $N_p \sim 15$. This indicates that horizon absorption may be observable at moderate SNRs in highly spinning unequal-mass systems with long inspirals, while for near-equal-mass binaries its detection is limited to very high-SNR events.

Since the spins have been held fixed in this analysis, it is useful to note how these results depend on them. As $\Delta\lambda_H \propto \tilde{\kappa}_H$, Fig.~\ref{fig:MM_absdeltalambda_3ifos_N_1024} implies that the mismatches above scale approximately as $\overline{\mathcal{MM}} \propto \tilde{\kappa}_H^2$, so that for the more typical $\ord{1}$ values of $|\tilde{\kappa}_H|$ of Fig.~\ref{fig:p_kH_q_0_1_chi_0_1_N_1e+10} they could be more than an order of magnitude smaller, correspondingly increasing the distinguishable SNRs.

\subsubsection{Parameter estimation studies}
\label{sec:Impact:Measure:PE}

Bayesian parameter estimation (PE) is the primary method used to analyze compact binary coalescence signals observed by GW detectors~\cite{Veitch:2014wba,Thrane:2018qnx}. While the mismatches studied in Sec.~\ref{sec:Impact:Measure:Mismatch} provide a rough indication of when signals with and without horizon absorption may become distinguishable, they are not sufficient to determine whether neglecting horizon absorption leads to biased parameter recovery or whether this effect can be detected in practice.

In Bayesian inference, we aim to compute the posterior probability distribution, $p(\bm{\theta}|d)$, of the signal parameters $\bm{\theta}$ given the detector data $d$ using Bayes' theorem,

\begin{equation}
    p(\bm{\theta}|d) = \frac{\mathcal{L}(d|\bm{\theta}) \pi(\bm{\theta})}{\mathcal{Z}} \, ,
    \label{eq:BayesTheorem}
\end{equation}

\noindent where $\pi(\bm{\theta})$ is the prior, $\mathcal{L}(d|\bm{\theta})$ is the likelihood, and $\mathcal{Z}$ is the evidence. Assuming stationary Gaussian noise, the likelihood reduces to the Whittle likelihood~\cite{Veitch:2014wba,Thrane:2018qnx},

\begin{align}
    \mathcal{L}(d|\bm{\theta}) & \propto \exp{\left\{ -\frac{1}{2} \sum_{i=1}^N  \langle h_i(\bm{\theta}) - d_i | h_i(\bm{\theta}) - d_i \rangle_i \right\}} \nonumber \\
    & \propto \exp{\left\{  \sum_{i=1}^N\left( \langle h_i(\bm{\theta}) | d_i \rangle_i -\frac{1}{2} \langle h_i(\bm{\theta}) | h_i(\bm{\theta}) \rangle_i \right) \right\}} \, ,
    \label{eq:Likelihood_def}    
\end{align}

\noindent where $d_i$ and $h_i$ represent the measured data (including noise) and the GW signal in the $i$-th detector, respectively. Similarly, $\langle \cdot \mid \cdot \rangle_i$ denotes the noise-weighted inner product introduced in Eq.~\eqref{eq:InnerProdDef}, evaluated using the PSD of the $i$-th detector. The evidence introduced in Eq.~\eqref{eq:BayesTheorem},

\begin{equation}
    \mathcal{Z} = \int \mathcal{L}(d|\bm{\theta}) \pi(\bm{\theta}) \d\bm{\theta} \, ,
    \label{eq:Evidence_def}
\end{equation}

\noindent ensures that the posterior probability distribution is normalized. The evidence measures the average likelihood over the prior volume and it is used for Bayesian model comparison~\cite{Kass:1995bf}. In particular, ratios of evidences between different hypotheses define Bayes factors, which quantify the degree to which the data prefer one model over another.

Due to the complexity of the integral in Eq.~\eqref{eq:Evidence_def}, stochastic sampling methods are required to explore the posterior distribution and estimate the evidence. Here, we perform PE using \texttt{bilby}~\cite{Ashton:2018jfp,Smith:2019ucc,Romero-Shaw:2020owr}, employing its implementation of the \texttt{dynesty}~\cite{Speagle:2020dqf} nested sampling algorithm. We perform zero-noise injections~\cite{Rodriguez:2013oaa} to avoid statistical fluctuations associated with individual noise realizations, as well as the additional uncertainty present in analyses of real events, where the true source parameters are unknown. Synthetic signals are injected into the LIGO Hanford (H1), LIGO Livingston (L1), and Virgo (V1) detectors, assuming the LIGO A+ and Virgo AdV+ sensitivity curves projected for O5~\cite{KAGRA:2013rdx,ObservingScenariosPSDs}. The data are analyzed from a minimum frequency of $f_\mathrm{min}=20\,\mathrm{Hz}$, which is also taken as the initial frequency of the \pyEFPEHM waveform. Consequently, the initial eccentricity $e_0$ is defined at $20\,\mathrm{Hz}$.

Due to the computational cost of PE, the number of analyses that can be performed is limited. We therefore focus on a fiducial system with component masses $m_1 = 14\,M_\odot$ and $m_2 = 1.4\,M_\odot$, and with the same spin vectors used throughout this work, namely $\bm{\chi}_{1,0} = [-0.2, 0.3, 0.9]$ and $\bm{\chi}_{2,0} = [-0.4, -0.4, 0.8]$. We further fix the initial mean anomaly $\ell_0 = 4.3$, inclination $\iota_0 = \pi/3$, reference phase $\phi_0 = 1.2$, polarization angle $\psi = 0.6$, right ascension $\alpha = 1.0$, declination $\delta = -0.316$, and coalescence GPS time $t_c = 1262276684\,\mathrm{s}$. We consider both quasi-circular injections with $e_0 = 0$ and eccentric injections with $e_0 = 0.3$. For each case, signals are injected at luminosity distances $d_L = 500$, $200$, $100$, and $50\,\mathrm{Mpc}$. The corresponding SNRs are listed in Table~\ref{table:PE_injections}. In all cases, the injected signals include horizon absorption, while the recovery is performed both with and without horizon absorption.

The secondary should be understood here as a low-mass black hole rather than a neutron star, although its nature is largely irrelevant, since as discussed in Sec.~\ref{sec:PNComputations:orbs}, horizon absorption is dominated by the primary, with $\kappa_{H,2}/\kappa_{H,1} \approx 8 \times 10^{-4}$ for this system, so that removing the horizon absorption of the secondary changes $\tilde{\kappa}_H$ by less than $0.1\%$.

\begin{table}[t]
    \centering
    \begin{tabular}{c|ccc}
        & $d_L\,[\mathrm{Mpc}]$ & Injection SNR & $\log{\mathcal{B}^\mathrm{HA}_\mathrm{no\,HA}}$\\
        \hline
        \multirow{4}{*}{$e_0 = 0$}
        & $500$ & 21.6 &  $-0.04 \pm 0.31$ \\
        & $200$ & 54.0 &  $0.06 \pm 0.36$ \\
        & $100$ & 108.0 &  $0.12 \pm 0.39$ \\
        & $50$  & 216.0 &  $0.32 \pm 0.43$ \\
        \hline
        \multirow{4}{*}{$e_0 = 0.3$}
        & $500$ & 21.5 & $0.13 \pm 0.34$ \\
        & $200$ & 53.8 & $1.19 \pm 0.39$ \\
        & $100$ & 107.6 & $3.02 \pm 0.42$ \\
        & $50$  & 215.3 & $8.60 \pm 0.45$ \\
        \hline
    \end{tabular}
    \caption{\justifying Summary of the injections including horizon absorption used in the PE analyses, labeled by their eccentricity and luminosity distance. For each injection, we report the optimal network SNR together with the logarithm of the Bayes factor comparing analyses performed with and without horizon absorption (Eq.~\eqref{eq:logB_HAnoHA}). Positive values of $\log{\mathcal{B}^\mathrm{HA}_\mathrm{no\,HA}}$ indicate a preference for the analyses including horizon absorption. The uncertainties in the log-Bayes factors are obtained by propagating the nested-sampling uncertainties in the evidences entering Eq.~\eqref{eq:logB_HAnoHA}.}
    \label{table:PE_injections}
\end{table}

To facilitate direct comparisons between analyses, identical priors are adopted for all PE runs. In particular, we use broad, uninformative priors that are uniform in the component masses, spin magnitudes, eccentricity, and coalescence time, and isotropic in sky position as well as binary and spin orientations. The luminosity distance prior assumes sources are distributed uniformly in comoving volume, using the \textit{Planck15} cosmology~\cite{Planck:2015fie}.

\begin{figure}[t!]
\centering  
\includegraphics[width=0.49\textwidth]{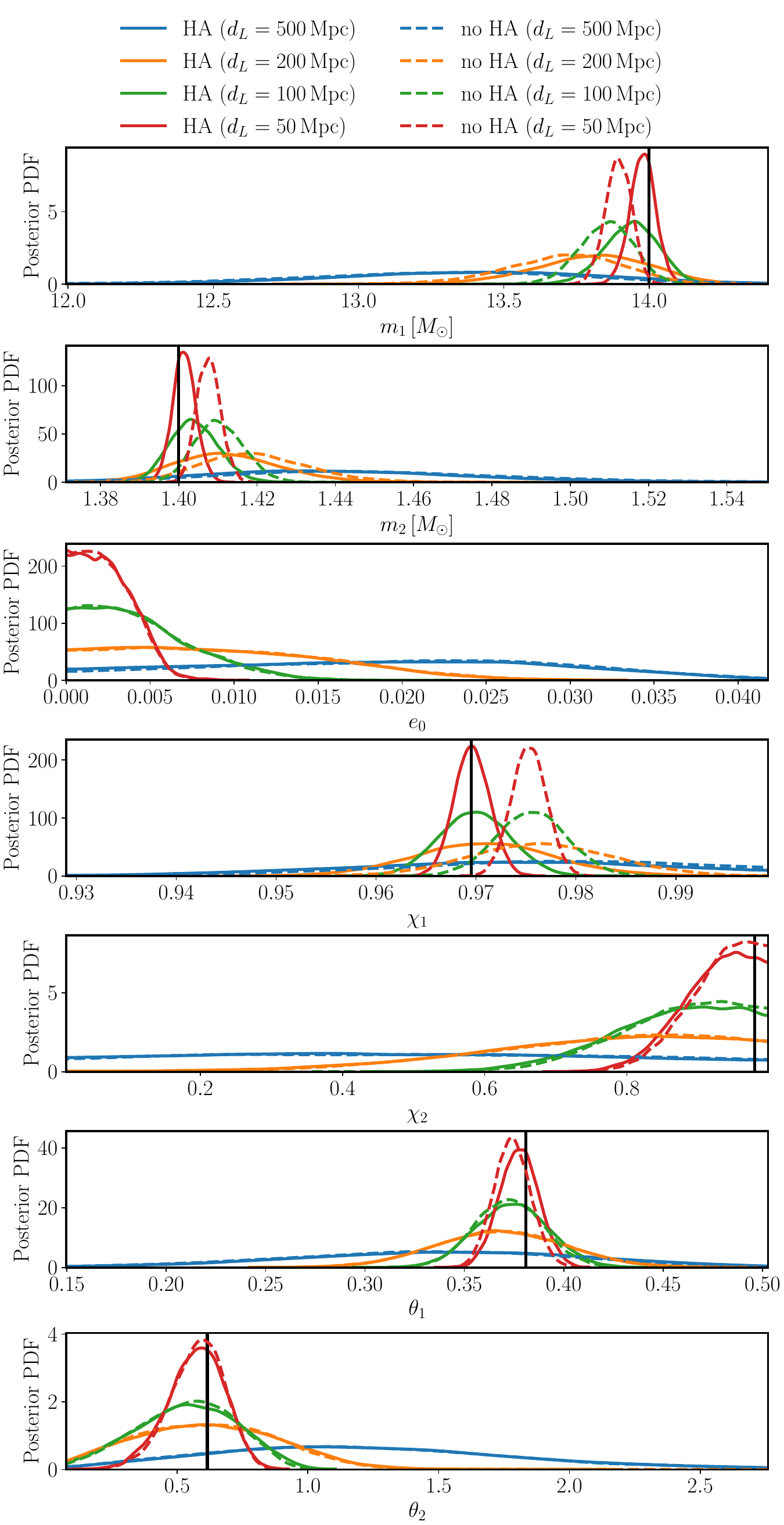}
\caption{\justifying Posterior probability density functions (PDFs) for the most relevant intrinsic parameters, obtained by the PE analyses of the quasi-circular ($e_0 = 0$) injections including horizon absorption. The analyses with horizon absorption (labeled ``HA'') are shown with solid lines, while the ones without (labeled ``no HA'') are shown with dashed lines. Different colors are used to represent the different luminosity distance of the injections. In each panel we show the posterior PDF for a different parameter. From top to bottom, we have the component masses $m_1$ and $m_2$, initial eccentricity $e_0$, the component dimensionless spin magnitudes $\chi_1$ and $\chi_2$ and tilt angles $\theta_1$ and $\theta_2$. 
}
\label{fig:densities_Inj_QCHA_32s_physically_important}
\end{figure}

\begin{figure}[t!]
\centering  
\includegraphics[width=0.49\textwidth]{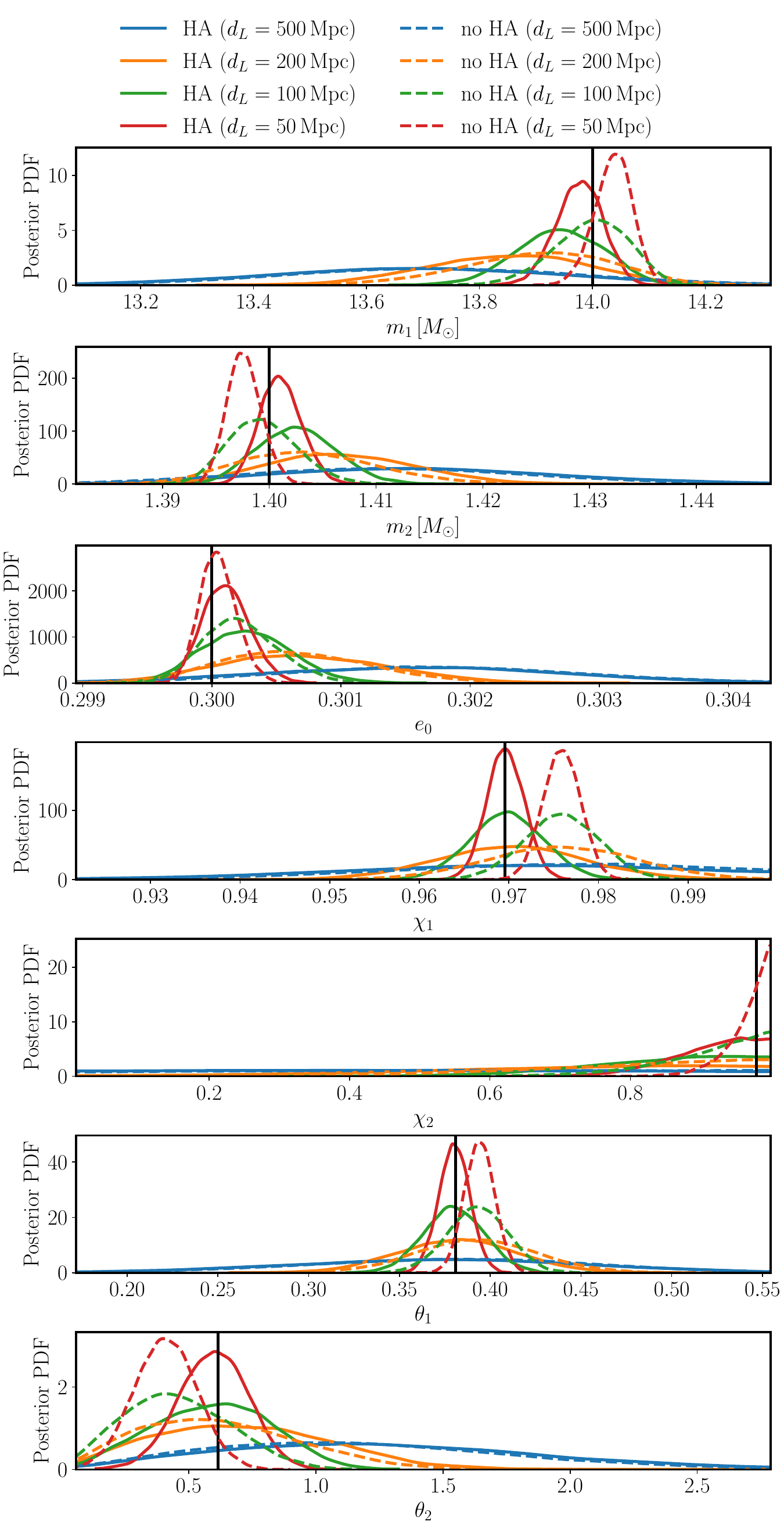}
\caption{\justifying Posterior probability density functions (PDFs) for the most relevant intrinsic parameters, obtained by the PE analyses of the eccentric ($e_0 = 0.3$) injections including horizon absorption. The analyses with horizon absorption (labeled ``HA'') are shown with solid lines, while the ones without (labeled ``no HA'') are shown with dashed lines. Different colors are used to represent the different luminosity distance of the injections. In each panel we show the posterior PDF for a different parameter. From top to bottom, we have the component masses $m_1$ and $m_2$, initial eccentricity $e_0$, the component dimensionless spin magnitudes $\chi_1$ and $\chi_2$ and tilt angles $\theta_1$ and $\theta_2$. 
}
\label{fig:densities_Inj_eccHA_32s_physically_important}
\end{figure}

In Figs.~\ref{fig:densities_Inj_QCHA_32s_physically_important} and~\ref{fig:densities_Inj_eccHA_32s_physically_important} we show the posterior probability density functions (PDFs) for the most relevant intrinsic parameters obtained from the analyses of the quasi-circular ($e_0 = 0$) and eccentric ($e_0 = 0.3$) injections, respectively. The minimized mismatch (Eq.~\ref{eq:MinMisMatchDef}) between waveforms with and without horizon absorption, evaluated at the injected parameters, is $0.017$ and $0.014$ for the quasi-circular and eccentric cases, respectively. Assuming an effective number of parameters of $N_p \sim 15$, these correspond to distinguishable SNRs of approximately $21$ and $23$.

The injections at $d_L = 500\,\mathrm{Mpc}$ have $\mathrm{SNR} \approx 22$ in both cases, placing them near the distinguishability threshold, and no clear differences are visible between the posteriors obtained with and without horizon absorption. However, for the injections at $d_L = 200\,\mathrm{Mpc}$ ($\mathrm{SNR} \approx 54$), small differences begin to appear in the posterior of the primary spin magnitude. For the highest-SNR injections, at $d_L = 100\,\mathrm{Mpc}$ ($\mathrm{SNR} \approx 108$) and $d_L = 50\,\mathrm{Mpc}$ ($\mathrm{SNR} \approx 220$), posterior differences become visible in multiple parameters and grow increasingly significant. While analyses including horizon absorption accurately recover the injected parameters, analyses neglecting it recover biased values for the component masses and the primary spin magnitude, and, in the eccentric case, also for the spin tilt angles.

\begin{figure}[t!]
\centering  
\includegraphics[width=0.49\textwidth]{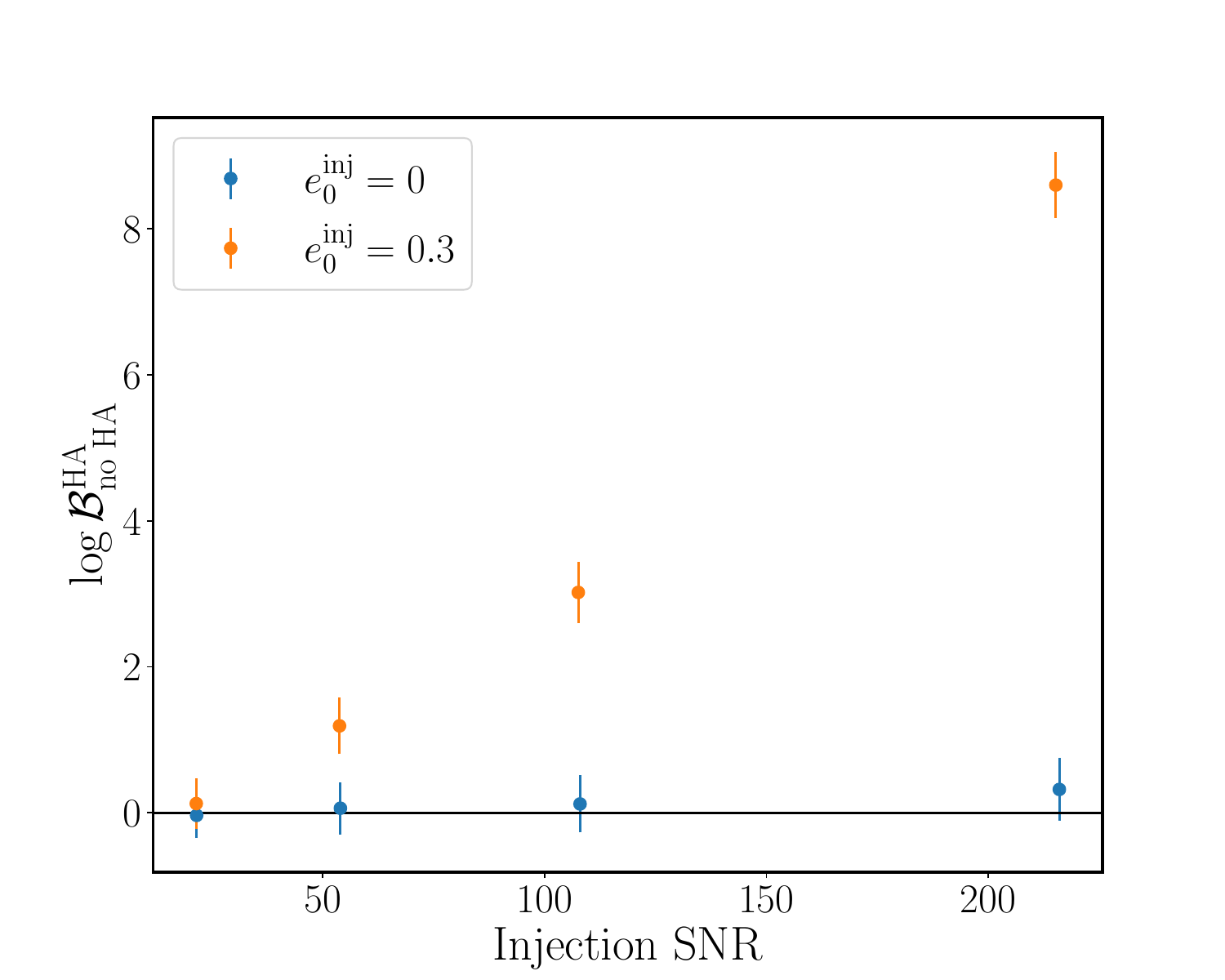}
\caption{\justifying Logarithm of the Bayes factor comparing analyses performed with and without horizon absorption (Eq.~\eqref{eq:logB_HAnoHA}) as a function of the SNR for the quasi-circular and eccentric injections. Positive values of $\log{\mathcal{B}^\mathrm{HA}_\mathrm{no\,HA}}$ indicate a preference for the analyses including horizon absorption. The error bars correspond to the uncertainty in the log-Bayes factor obtained through propagation of the nested-sampling uncertainties in the evidences entering Eq.~\eqref{eq:logB_HAnoHA}.
}
\label{fig:lnB_HAnoHA}
\end{figure}

The biases in the recovered parameters found in Figs.~\ref{fig:densities_Inj_QCHA_32s_physically_important} and~\ref{fig:densities_Inj_eccHA_32s_physically_important} when neglecting horizon absorption can affect the inferred astrophysical interpretation of the events and, potentially, the estimated properties of the compact binary population. Nonetheless, parameter biases alone do not provide definitive evidence for the presence or absence of horizon absorption in the signal, since there is no independent measurement of the true binary parameters against which the recovered values can be compared. To assess whether the data statistically favor waveform models including horizon absorption, we instead consider the Bayes factors,

\begin{equation}
    \log{\mathcal{B}^\mathrm{HA}_\mathrm{no\,HA}} = \log{\frac{\mathcal{Z}_\mathrm{HA}}{\mathcal{Z}_\mathrm{no\,HA}}} \, ,
    \label{eq:logB_HAnoHA}
\end{equation}

\noindent where $\mathcal{Z}_\mathrm{HA}$ and $\mathcal{Z}_\mathrm{no\,HA}$ are the evidences obtained from the PE analyses with and without horizon absorption, respectively, and $\log$ denotes the natural logarithm, as throughout this work.

The resulting Bayes factors are reported in Table~\ref{table:PE_injections} and shown as a function of SNR in Fig.~\ref{fig:lnB_HAnoHA}. For the quasi-circular injections, the Bayes factors remain consistent with zero even for the loudest injections, indicating that the effect of horizon absorption can be almost fully absorbed by shifts in the binary parameters. In contrast, for the eccentric injections, the Bayes factor increases steadily with SNR and reaches $\log{\mathcal{B}^\mathrm{HA}_\mathrm{no\,HA}} \approx 9$ for the loudest case. This suggests that, in the eccentric case where the signal morphology is more complex, shifts in the parameters of the waveform without horizon absorption are less able to reproduce the signal including horizon absorption, making the presence of this effect more readily detectable.

\section{Conclusion}
\label{sec:conclusion}

In this work, we have derived for the first time the effect of horizon absorption in eccentric, precessing binary black hole inspirals. In particular, we obtained the leading-order correction for spinning black holes, which enters the binary evolution at 2.5PN order. We then incorporated these corrections into \pyEFPEHM, resulting in an eccentric, precessing inspiral waveform model that consistently includes horizon absorption effects.

We analytically studied the orbital dephasing induced by horizon absorption and compared it against the dephasing predicted by \pyEFPEHM, finding good agreement between the analytical estimates and the numerical waveform model. We find that the dephasing is largest for systems with large spin components aligned or anti-aligned with the orbital angular momentum ($|\bm{\chi}_i \cdot \uvec{l}| \sim 1$), highly unequal mass ratios ($q = m_2/m_1 \ll 1$), and long inspirals spanning a wide frequency range ($\log(y_f/y_0) \gg 1$).

We computed mismatches between \pyEFPEHM waveforms with and without horizon absorption for LIGO, ET, and LISA sources. We found that the mismatch is strongly correlated with the estimated dephasing of Eq.~\eqref{eq:Deltalambda_LO}, following the analytical expectation $\overline{\mathcal{MM}} \propto |\Delta\lambda_H|^2$~\cite{Lindblom:2008cm}, so that this dephasing provides an inexpensive proxy for the detectability of horizon absorption, even in eccentric, precessing binaries. Studying the dependence of the mismatch on the binary parameters, we obtained mismatches exceeding $1\%$ for highly spinning binaries with mass ratios $q \lesssim 0.2$, indicating that horizon absorption may become observable in realistic signals detected by current and future ground- and space-based GW detectors.

To investigate the detectability of these effects in more detail, we performed full Bayesian parameter estimation analyses on simulated signals injected into the LIGO--Virgo detector network, comparing waveform recoveries performed with and without horizon absorption. We found that neglecting horizon absorption leads to systematic parameter biases in both quasi-circular and eccentric systems, particularly in the primary spin magnitude and the component masses. For quasi-circular binaries, neglecting horizon absorption can be largely compensated by shifts in the intrinsic binary parameters, resulting in Bayes factors that remain consistent with no preference for models including horizon absorption, even at very high SNRs. In contrast, for eccentric binaries, where the waveform morphology is more complex, parameter shifts are less effective at reproducing signals that include horizon absorption when this effect is neglected in the recovery. As a result, the Bayes factor increasingly favors waveform models including horizon absorption as the SNR grows, making the effect potentially detectable in high-SNR eccentric events.

We note that in current waveform models calibrated against NR simulations, such as the SEOBNR~\cite{Pompili:2023tna,Gamboa:2026jht} and IMRPhenom~\cite{Pratten:2020fqn,Estelles:2020twz} families, horizon absorption is partly reabsorbed into the calibration even when not modeled explicitly. As our PE results suggest, however, this compensation might be less effective for eccentric, precessing binaries, where the richer signal morphology and more complicated parameter dependence of the horizon absorption correction make the effect harder to mimic.

In this work we have only derived the leading-order correction due to horizon absorption. This contribution is expected to usually dominate, since entering at 2.5PN means it accumulates logarithmically over long inspirals. Although higher-order corrections entering at 3.5PN order and beyond do not accumulate over long inspirals, they can still become relevant for highly unequal-mass-ratio systems, as their contribution to the dephasing also scales as $1/\nu$. Note that for sufficiently small black hole spins, the non-spinning horizon absorption correction that first appears at 4PN can dominate over the leading-order spinning correction studied here. For aligned-spin binaries, horizon absorption corrections are currently known up to 4PN order~\cite{Chiaramello:2024unv}, and extending these results to generic eccentric and precessing systems would therefore be an important next step. 

In the extreme-mass-ratio regime where such higher-order PN contributions become important, the dynamics is expected to be more accurately captured by gravitational self-force (SF) methods rather than a purely PN description. Nevertheless, these higher-order PN corrections can still play a relevant role in the construction of accurate SF-PN hybrid waveform models~\cite{Honet:2025gge,Honet:2025lmk}.

\section*{Code Availability}

The repository containing the waveform model, along with scripts to reproduce the figures in this paper, will be made available at Ref.~\cite{pyEFPEHM_repo}. The Bayesian optimization algorithm used for mismatch minimization is publicly available at Ref.~\cite{pybop_repo}.

\section*{Acknowledgments}

We thank Juan Garc\'ia-Bellido and Diganta Bandopadhyay for helpful discussion.
We thank Rosella Gamba for her helpful comments and suggestions as internal LIGO reviewer.
GM's research is supported in part by the European Research Council (ERC) Horizon Synergy Grant “Making Sense of the Unexpected in the Gravitational-Wave Sky” grant agreement no. GWSky–101167314.
We acknowledge the computational resources provided by the Max Planck Institute for Gravitational Physics (Albert Einstein Institute), Potsdam, in particular, the Hypatia cluster.
This material is based upon work supported by NSF's LIGO Laboratory which is a major facility fully funded by the National Science Foundation.
This manuscript has LIGO document number P2600288.

\bibliography{Refs}

@article{Poisson:2004cw,
    author = "Poisson, Eric",
    title = "{Absorption of mass and angular momentum by a black hole: Time-domain formalisms for gravitational perturbations, and the small-hole / slow-motion approximation}",
    eprint = "gr-qc/0407050",
    archivePrefix = "arXiv",
    doi = "10.1103/PhysRevD.70.084044",
    journal = "Phys. Rev. D",
    volume = "70",
    pages = "084044",
    year = "2004"
}

@article{Chiaramello:2024unv,
    author = "Chiaramello, Danilo and Gamba, Rossella",
    title = "{Horizon absorption on noncircular, planar binary black hole dynamics}",
    eprint = "2408.15322",
    archivePrefix = "arXiv",
    primaryClass = "gr-qc",
    doi = "10.1103/PhysRevD.111.024024",
    journal = "Phys. Rev. D",
    volume = "111",
    number = "2",
    pages = "024024",
    year = "2025"
}

@article{Saketh:2022xjb,
    author = "Saketh, M. V. S. and Steinhoff, Jan and Vines, Justin and Buonanno, Alessandra",
    title = "{Modeling horizon absorption in spinning binary black holes using effective worldline theory}",
    eprint = "2212.13095",
    archivePrefix = "arXiv",
    primaryClass = "gr-qc",
    doi = "10.1103/PhysRevD.107.084006",
    journal = "Phys. Rev. D",
    volume = "107",
    number = "8",
    pages = "084006",
    year = "2023"
}

@article{Thorne:1984mz,
    author = "Thorne, Kip S. and Hartle, James B.",
    title = "{Laws of motion and precession for black holes and other bodies}",
    doi = "10.1103/PhysRevD.31.1815",
    journal = "Phys. Rev. D",
    volume = "31",
    pages = "1815--1837",
    year = "1984"
}

@article{Chia:2020yla,
    author = "Chia, Horng Sheng",
    title = "{Tidal deformation and dissipation of rotating black holes}",
    eprint = "2010.07300",
    archivePrefix = "arXiv",
    primaryClass = "gr-qc",
    doi = "10.1103/PhysRevD.104.024013",
    journal = "Phys. Rev. D",
    volume = "104",
    number = "2",
    pages = "024013",
    year = "2021"
}

@article{Charalambous:2021mea,
    author = "Charalambous, Panagiotis and Dubovsky, Sergei and Ivanov, Mikhail M.",
    title = "{On the Vanishing of Love Numbers for Kerr Black Holes}",
    eprint = "2102.08917",
    archivePrefix = "arXiv",
    primaryClass = "hep-th",
    reportNumber = "INR-TH-2021-001",
    doi = "10.1007/JHEP05(2021)038",
    journal = "JHEP",
    volume = "05",
    pages = "038",
    year = "2021"
}

@article{Datta:2023wsn,
    author = "Datta, Sayak",
    title = "{Horizon fluxes of binary black holes in eccentric orbits}",
    eprint = "2305.03771",
    archivePrefix = "arXiv",
    primaryClass = "gr-qc",
    doi = "10.1140/epjc/s10052-024-13371-8",
    journal = "Eur. Phys. J. C",
    volume = "84",
    number = "10",
    pages = "1077",
    year = "2024"
}

@article{Taylor:2008xy,
    author = "Taylor, Stephanne and Poisson, Eric",
    title = "{Nonrotating black hole in a post-Newtonian tidal environment}",
    eprint = "0806.3052",
    archivePrefix = "arXiv",
    primaryClass = "gr-qc",
    doi = "10.1103/PhysRevD.78.084016",
    journal = "Phys. Rev. D",
    volume = "78",
    pages = "084016",
    year = "2008"
}

@article{Alvi:2001mx,
    author = "Alvi, Kashif",
    title = "{Energy and angular momentum flow into a black hole in a binary}",
    eprint = "gr-qc/0107080",
    archivePrefix = "arXiv",
    doi = "10.1103/PhysRevD.64.104020",
    journal = "Phys. Rev. D",
    volume = "64",
    pages = "104020",
    year = "2001"
}

@article{Chatziioannou:2012gq,
    author = "Chatziioannou, Katerina and Poisson, Eric and Yunes, Nicolas",
    title = "{Tidal heating and torquing of a Kerr black hole to next-to-leading order in the tidal coupling}",
    eprint = "1211.1686",
    archivePrefix = "arXiv",
    primaryClass = "gr-qc",
    doi = "10.1103/PhysRevD.87.044022",
    journal = "Phys. Rev. D",
    volume = "87",
    number = "4",
    pages = "044022",
    year = "2013"
}

@article{Henry:2023tka,
    author = "Henry, Quentin and Khalil, Mohammed",
    title = "{Spin effects in gravitational waveforms and fluxes for binaries on eccentric orbits to the third post-Newtonian order}",
    eprint = "2308.13606",
    archivePrefix = "arXiv",
    primaryClass = "gr-qc",
    doi = "10.1103/PhysRevD.108.104016",
    journal = "Phys. Rev. D",
    volume = "108",
    number = "10",
    pages = "104016",
    year = "2023"
}

@article{Klein:2010ti,
    author = "Klein, Antoine and Jetzer, Philippe",
    title = "{Spin effects in the phasing of gravitational waves from binaries on eccentric orbits}",
    eprint = "1005.2046",
    archivePrefix = "arXiv",
    primaryClass = "gr-qc",
    doi = "10.1103/PhysRevD.81.124001",
    journal = "Phys. Rev. D",
    volume = "81",
    pages = "124001",
    year = "2010"
}

@article{Racine:2008qv,
    author = "Racine, Etienne",
    title = "{Analysis of spin precession in binary black hole systems including quadrupole-monopole interaction}",
    eprint = "0803.1820",
    archivePrefix = "arXiv",
    primaryClass = "gr-qc",
    doi = "10.1103/PhysRevD.78.044021",
    journal = "Phys. Rev. D",
    volume = "78",
    pages = "044021",
    year = "2008"
}

@article{Barker:1979jmv,
    author = "Barker, B. M. and O'Connell, R. F.",
    title = "{The gravitational interaction: Spin, rotation, and quantum effects-a review}",
    doi = "10.1007/BF00756587",
    journal = "Gen. Rel. Grav.",
    volume = "11",
    number = "2",
    pages = "149--175",
    year = "1979"
}

@article{Klein:2018ybm,
    author = "Klein, Antoine and Boetzel, Yannick and Gopakumar, Achamveedu and Jetzer, Philippe and de Vittori, Lorenzo",
    title = "{Fourier domain gravitational waveforms for precessing eccentric binaries}",
    eprint = "1801.08542",
    archivePrefix = "arXiv",
    primaryClass = "gr-qc",
    doi = "10.1103/PhysRevD.98.104043",
    journal = "Phys. Rev. D",
    volume = "98",
    number = "10",
    pages = "104043",
    year = "2018"
}

@article{Klein:2021jtd,
    author = "Klein, Antoine",
    title = "{EFPE: Efficient fully precessing eccentric gravitational waveforms for binaries with long inspirals}",
    eprint = "2106.10291",
    archivePrefix = "arXiv",
    primaryClass = "gr-qc",
    month = "6",
    year = "2021",
    journal = "",
}

@article{Bohe:2012mr,
    author = "Bohe, Alejandro and Marsat, Sylvain and Faye, Guillaume and Blanchet, Luc",
    title = "{Next-to-next-to-leading order spin-orbit effects in the near-zone metric and precession equations of compact binaries}",
    eprint = "1212.5520",
    archivePrefix = "arXiv",
    primaryClass = "gr-qc",
    doi = "10.1088/0264-9381/30/7/075017",
    journal = "Class. Quant. Grav.",
    volume = "30",
    pages = "075017",
    year = "2013"
}

@misc{Sturani:2015STA,
  author       = "Sturani, Riccardo",
  title        = "{Note on the derivation of the angular momentum and spin precessing equations in SpinTaylor codes}",
  url          = {https://dcc.ligo.org/public/0122/T1500554/013/dLdS.pdf},
  year         = "2015"
}

@article{Akcay:2020qrj,
    author = "Akcay, Sarp and Gamba, Rossella and Bernuzzi, Sebastiano",
    title = "{Hybrid post-Newtonian effective-one-body scheme for spin-precessing compact-binary waveforms up to merger}",
    eprint = "2005.05338",
    archivePrefix = "arXiv",
    primaryClass = "gr-qc",
    doi = "10.1103/PhysRevD.103.024014",
    journal = "Phys. Rev. D",
    volume = "103",
    number = "2",
    pages = "024014",
    year = "2021"
}

@article{Khalil:2023kep,
    author = "Khalil, Mohammed and Buonanno, Alessandra and Estelles, Hector and Mihaylov, Deyan P. and Ossokine, Serguei and Pompili, Lorenzo and Ramos-Buades, Antoni",
    title = "{Theoretical groundwork supporting the precessing-spin two-body dynamics of the effective-one-body waveform models SEOBNRv5}",
    eprint = "2303.18143",
    archivePrefix = "arXiv",
    primaryClass = "gr-qc",
    doi = "10.1103/PhysRevD.108.124036",
    journal = "Phys. Rev. D",
    volume = "108",
    number = "12",
    pages = "124036",
    year = "2023"
}

@article{Morras:2025nlp,
    author = "Morras, Gonzalo and Pratten, Geraint and Schmidt, Patricia",
    title = "{Improved post-Newtonian waveform model for inspiralling precessing-eccentric compact binaries}",
    eprint = "2502.03929",
    archivePrefix = "arXiv",
    primaryClass = "gr-qc",
    reportNumber = "IFT-UAM/CSIC-25-12",
    doi = "10.1103/PhysRevD.111.084052",
    journal = "Phys. Rev. D",
    volume = "111",
    number = "8",
    pages = "084052",
    year = "2025"
}

@article{Boetzel:2017zza,
    author = "Boetzel, Yannick and Susobhanan, Abhimanyu and Gopakumar, Achamveedu and Klein, Antoine and Jetzer, Philippe",
    title = "{Solving post-Newtonian accurate Kepler Equation}",
    eprint = "1707.02088",
    archivePrefix = "arXiv",
    primaryClass = "gr-qc",
    doi = "10.1103/PhysRevD.96.044011",
    journal = "Phys. Rev. D",
    volume = "96",
    number = "4",
    pages = "044011",
    year = "2017"
}

@article{Lindblom:2008cm,
    author = "Lindblom, Lee and Owen, Benjamin J. and Brown, Duncan A.",
    title = "{Model Waveform Accuracy Standards for Gravitational Wave Data Analysis}",
    eprint = "0809.3844",
    archivePrefix = "arXiv",
    primaryClass = "gr-qc",
    doi = "10.1103/PhysRevD.78.124020",
    journal = "Phys. Rev. D",
    volume = "78",
    pages = "124020",
    year = "2008"
}

@article{Thompson:2025hhc,
    author = "Thompson, Jonathan E. and Hoy, Charlie and Fauchon-Jones, Edward and Hannam, Mark",
    title = "{Use and interpretation of signal-model indistinguishability measures for gravitational-wave astronomy}",
    eprint = "2506.10530",
    archivePrefix = "arXiv",
    primaryClass = "gr-qc",
    reportNumber = "LIGO-P2500361",
    doi = "10.1103/ddz7-x9zz",
    journal = "Phys. Rev. D",
    volume = "112",
    number = "6",
    pages = "064011",
    year = "2025"
}

@article{Schutz:1987xok,
    author = "Schutz, Bernard F. and Tinto, Massimo",
    title = "{Antenna patterns of interferometric detectors of gravitational waves {\textendash} I. Linearly polarized waves}",
    doi = "10.1093/mnras/224.1.131",
    journal = "Mon. Not. Roy. Astron. Soc.",
    volume = "224",
    number = "1",
    pages = "131--154",
    year = "1987"
}

@article{Finn:1992xs,
    author = "Finn, Lee Samuel and Chernoff, David F.",
    title = "{Observing binary inspiral in gravitational radiation: One interferometer}",
    eprint = "gr-qc/9301003",
    archivePrefix = "arXiv",
    reportNumber = "PRINT-93-0138 (NORTHWESTERN)",
    doi = "10.1103/PhysRevD.47.2198",
    journal = "Phys. Rev. D",
    volume = "47",
    pages = "2198--2219",
    year = "1993"
}

@article{Pompili:2023tna,
    author = "Pompili, Lorenzo and others",
    title = "{Laying the foundation of the effective-one-body waveform models SEOBNRv5: Improved accuracy and efficiency for spinning nonprecessing binary black holes}",
    eprint = "2303.18039",
    archivePrefix = "arXiv",
    primaryClass = "gr-qc",
    doi = "10.1103/PhysRevD.108.124035",
    journal = "Phys. Rev. D",
    volume = "108",
    number = "12",
    pages = "124035",
    year = "2023"
}

@article{Harry:2016ijz,
    author = "Harry, Ian and Privitera, Stephen and Boh\'e, Alejandro and Buonanno, Alessandra",
    title = "{Searching for Gravitational Waves from Compact Binaries with Precessing Spins}",
    eprint = "1603.02444",
    archivePrefix = "arXiv",
    primaryClass = "gr-qc",
    doi = "10.1103/PhysRevD.94.024012",
    journal = "Phys. Rev. D",
    volume = "94",
    number = "2",
    pages = "024012",
    year = "2016"
}

@article{Sathyaprakash:1991mt,
    author = "Sathyaprakash, B. S. and Dhurandhar, S. V.",
    title = "{Choice of filters for the detection of gravitational waves from coalescing binaries}",
    doi = "10.1103/PhysRevD.44.3819",
    journal = "Phys. Rev. D",
    volume = "44",
    pages = "3819--3834",
    year = "1991"
}

@article{Brent:1971,
    author = {Brent, R. P.},
    title = {An algorithm with guaranteed convergence for finding a zero of a function},
    journal = {The Computer Journal},
    volume = {14},
    number = {4},
    pages = {422-425},
    year = {1971},
    month = {01},
    issn = {0010-4620},
    doi = {10.1093/comjnl/14.4.422},
    url = {https://doi.org/10.1093/comjnl/14.4.422},
    eprint = {https://academic.oup.com/comjnl/article-pdf/14/4/422/927778/140422.pdf},
}

@article{KAGRA:2013rdx,
    author = "Abbott, B. P. and others",
    collaboration = "KAGRA, LIGO Scientific, Virgo",
    title = "{Prospects for observing and localizing gravitational-wave transients with Advanced LIGO, Advanced Virgo and KAGRA}",
    eprint = "1304.0670",
    archivePrefix = "arXiv",
    primaryClass = "gr-qc",
    reportNumber = "LIGO-P1200087, VIR-0288A-12, JGW-P1808427",
    doi = "10.1007/s41114-020-00026-9",
    journal = "Living Rev. Rel.",
    volume = "19",
    pages = "1",
    year = "2016"
}

@techreport{ObservingScenariosPSDs,
      author  = "B. P. Abbott and others",
      title   = "{Noise curves used for Simulations in the update of the Observing Scenarios Paper}",
      number  = "LIGO-T2000012",
      institution = "{LIGO Virgo KAGRA Collaboration}",
      url     = "https://dcc.ligo.org/LIGO-T2000012/public",
      year    = "2020"
}

@techreport{ET_PSDs_CoBA,
  title        = "{ET sensitivity curves used for CoBA Science Study}",
  author       = "Danilishin, Stefan and Zhang, Teng",
  institution  = "Einstein Telescope Collaboration",
  year         = "2023",
  number       = "ET-0304B-22",
  url          = "https://apps.et-gw.eu/tds/ql/?c=16492",
}

@article{Robson:2018ifk,
    author = "Robson, Travis and Cornish, Neil J. and Liu, Chang",
    title = "{The construction and use of LISA sensitivity curves}",
    eprint = "1803.01944",
    archivePrefix = "arXiv",
    primaryClass = "astro-ph.HE",
    doi = "10.1088/1361-6382/ab1101",
    journal = "Class. Quant. Grav.",
    volume = "36",
    number = "10",
    pages = "105011",
    year = "2019"
}

@article{Peters:1964zz,
    author = "Peters, P. C.",
    title = "{Gravitational Radiation and the Motion of Two Point Masses}",
    doi = "10.1103/PhysRev.136.B1224",
    journal = "Phys. Rev.",
    volume = "136",
    pages = "B1224--B1232",
    year = "1964"
}

@article{Veitch:2014wba,
    author = "Veitch, J. and others",
    title = "{Parameter estimation for compact binaries with ground-based gravitational-wave observations using the LALInference software library}",
    eprint = "1409.7215",
    archivePrefix = "arXiv",
    primaryClass = "gr-qc",
    reportNumber = "LIGO-P1400152",
    doi = "10.1103/PhysRevD.91.042003",
    journal = "Phys. Rev. D",
    volume = "91",
    number = "4",
    pages = "042003",
    year = "2015"
}

@article{Thrane:2018qnx,
    author = "Thrane, Eric and Talbot, Colm",
    title = "{An introduction to Bayesian inference in gravitational-wave astronomy: parameter estimation, model selection, and hierarchical models}",
    eprint = "1809.02293",
    archivePrefix = "arXiv",
    primaryClass = "astro-ph.IM",
    doi = "10.1017/pasa.2019.2",
    journal = "Publ. Astron. Soc. Austral.",
    volume = "36",
    pages = "e010",
    year = "2019",
    note = "[Erratum: Publ.Astron.Soc.Austral. 37, e036 (2020)]"
}

@article{Ashton:2018jfp,
    author = "Ashton, Gregory and others",
    title = "{BILBY: A user-friendly Bayesian inference library for gravitational-wave astronomy}",
    eprint = "1811.02042",
    archivePrefix = "arXiv",
    primaryClass = "astro-ph.IM",
    doi = "10.3847/1538-4365/ab06fc",
    journal = "Astrophys. J. Suppl.",
    volume = "241",
    number = "2",
    pages = "27",
    year = "2019"
}

@article{Speagle:2020dqf,
    author = "Speagle, Joshua S.",
    title = "{DYNESTY: a dynamic nested sampling package for estimating Bayesian posteriors and evidences}",
    eprint = "1904.02180",
    archivePrefix = "arXiv",
    primaryClass = "astro-ph.IM",
    doi = "10.1093/mnras/staa278",
    journal = "Mon. Not. Roy. Astron. Soc.",
    volume = "493",
    number = "3",
    pages = "3132--3158",
    year = "2020",
    month = "04",
}

@article{Romero-Shaw:2020owr,
    author = "Romero-Shaw, I. M. and others",
    title = "{Bayesian inference for compact binary coalescences with bilby: validation and application to the first LIGO\textendash{}Virgo gravitational-wave transient catalogue}",
    eprint = "2006.00714",
    archivePrefix = "arXiv",
    primaryClass = "astro-ph.IM",
    doi = "10.1093/mnras/staa2850",
    journal = "Mon. Not. Roy. Astron. Soc.",
    volume = "499",
    number = "3",
    pages = "3295--3319",
    year = "2020"
}

@article{Smith:2019ucc,
    author = "Smith, Rory J. E. and Ashton, Gregory and Vajpeyi, Avi and Talbot, Colm",
    title = "{Massively parallel Bayesian inference for transient gravitational-wave astronomy}",
    eprint = "1909.11873",
    archivePrefix = "arXiv",
    primaryClass = "gr-qc",
    reportNumber = "LIGO Document P1900255-v1",
    doi = "10.1093/mnras/staa2483",
    journal = "Mon. Not. Roy. Astron. Soc.",
    volume = "498",
    number = "3",
    pages = "4492--4502",
    year = "2020"
}

@article{Rodriguez:2013oaa,
    author = "Rodriguez, Carl L. and Farr, Benjamin and Raymond, Vivien and Farr, Will M. and Littenberg, Tyson B. and Fazi, Diego and Kalogera, Vicky",
    title = "{Basic Parameter Estimation of Binary Neutron Star Systems by the Advanced LIGO/Virgo Network}",
    eprint = "1309.3273",
    archivePrefix = "arXiv",
    primaryClass = "astro-ph.HE",
    doi = "10.1088/0004-637X/784/2/119",
    journal = "Astrophys. J.",
    volume = "784",
    pages = "119",
    year = "2014"
}

@article{Planck:2015fie,
    author = "Ade, P. A. R. and others",
    collaboration = "Planck",
    title = "{Planck 2015 results. XIII. Cosmological parameters}",
    eprint = "1502.01589",
    archivePrefix = "arXiv",
    primaryClass = "astro-ph.CO",
    doi = "10.1051/0004-6361/201525830",
    journal = "Astron. Astrophys.",
    volume = "594",
    pages = "A13",
    year = "2016"
}

@article{Bautista:2024emt,
    author = "Bautista, Yilber Fabian and Huang, Yu-Tin and Kim, Jung-Wook",
    title = "{Absorptive effects in black hole scattering}",
    eprint = "2411.03382",
    archivePrefix = "arXiv",
    primaryClass = "hep-th",
    doi = "10.1103/PhysRevD.111.044043",
    journal = "Phys. Rev. D",
    volume = "111",
    number = "4",
    pages = "044043",
    year = "2025"
}

@article{Gamba:2026fqa,
    author = "Gamba, Rossella and Chiaramello, Danilo and Shukla, Estuti and Albanesi, Simone",
    title = "{Spin the black circle II: tidal heating and torquing of a rotating black hole by a test mass on generic orbits}",
    eprint = "2603.28982",
    archivePrefix = "arXiv",
    primaryClass = "gr-qc",
    month = "3",
    year = "2026",
    journal = ""
}

@article{Kass:1995bf,
  title={Bayes factors},
  author={Kass, Robert E and Raftery, Adrian E},
  journal={Journal of the american statistical association},
  volume={90},
  number={430},
  pages={773--795},
  year={1995},
  publisher={Taylor \& Francis}
}

@article{Morras:2026fho,
    author = "Morras, Gonzalo and Pratten, Geraint and Schmidt, Patricia and Buonanno, Alessandra",
    title = "{Post-Newtonian inspiral waveform model for eccentric precessing binaries with higher-order modes and matter effects}",
    eprint = "2604.11903",
    archivePrefix = "arXiv",
    primaryClass = "gr-qc",
    month = "4",
    year = "2026",
    journal = ""
}

@book{Colwell:1993book,
  author       = {P. Colwell},
  title        = {Solving Kepler's Equation over Three Centuries},
  publisher    = {Willmann-Bell},
  year         = {1993}
}

@article{Damour:1985ecc,
  author       = {T. Damour and N. Deruelle},
  title        = {General relativistic celestial mechanics of binary systems. I. The post-newtonian motion},
  journal      = {Ann. Inst. Henri Poincaré Phys. Théor.},
  volume       = {43},
  number       = {1},
  pages        = {107--132},
  year         = {1985}
}

@article{Mukherjee:2023pge,
    author = "Mukherjee, Samanwaya and Phukon, Khun Sang and Datta, Sayak and Bose, Sukanta",
    title = "{Phenomenological gravitational waveform model of binary black holes incorporating horizon fluxes}",
    eprint = "2311.17554",
    archivePrefix = "arXiv",
    primaryClass = "gr-qc",
    reportNumber = "LIGO-P2300183",
    doi = "10.1103/PhysRevD.110.124027",
    journal = "Phys. Rev. D",
    volume = "110",
    number = "12",
    pages = "124027",
    year = "2024"
}

@article{Mukherjee:2022wws,
    author = "Mukherjee, Samanwaya and Datta, Sayak and Tiwari, Srishti and Phukon, Khun Sang and Bose, Sukanta",
    title = "{Toward establishing the presence or absence of horizons in coalescing binaries of compact objects by using their gravitational wave signals}",
    eprint = "2202.08661",
    archivePrefix = "arXiv",
    primaryClass = "gr-qc",
    reportNumber = "LIGO-P2100474",
    doi = "10.1103/PhysRevD.106.104032",
    journal = "Phys. Rev. D",
    volume = "106",
    number = "10",
    pages = "104032",
    year = "2022"
}

@article{Datta:2019epe,
    author = "Datta, Sayak and Brito, Richard and Bose, Sukanta and Pani, Paolo and Hughes, Scott A.",
    title = "{Tidal heating as a discriminator for horizons in extreme mass ratio inspirals}",
    eprint = "1910.07841",
    archivePrefix = "arXiv",
    primaryClass = "gr-qc",
    doi = "10.1103/PhysRevD.101.044004",
    journal = "Phys. Rev. D",
    volume = "101",
    number = "4",
    pages = "044004",
    year = "2020"
}

@article{Datta:2024vll,
    author = "Datta, Sayak and Brito, Richard and Hughes, Scott A. and Klinger, Talya and Pani, Paolo",
    title = "{Tidal heating as a discriminator for horizons in equatorial eccentric extreme mass ratio inspirals}",
    eprint = "2404.04013",
    archivePrefix = "arXiv",
    primaryClass = "gr-qc",
    doi = "10.1103/PhysRevD.110.024048",
    journal = "Phys. Rev. D",
    volume = "110",
    number = "2",
    pages = "024048",
    year = "2024"
}

@article{LIGOScientific:2014pky,
    author = "Aasi, J. and others",
    collaboration = "LIGO Scientific",
    title = "{Advanced LIGO}",
    eprint = "1411.4547",
    archivePrefix = "arXiv",
    primaryClass = "gr-qc",
    doi = "10.1088/0264-9381/32/7/074001",
    journal = "Class. Quant. Grav.",
    volume = "32",
    pages = "074001",
    year = "2015"
}

@article{VIRGO:2014yos,
    author = "Acernese, F. and others",
    collaboration = "VIRGO",
    title = "{Advanced Virgo: a second-generation interferometric gravitational wave detector}",
    eprint = "1408.3978",
    archivePrefix = "arXiv",
    primaryClass = "gr-qc",
    doi = "10.1088/0264-9381/32/2/024001",
    journal = "Class. Quant. Grav.",
    volume = "32",
    number = "2",
    pages = "024001",
    year = "2015"
}

@article{KAGRA:2018plz,
    author = "Akutsu, T. and others",
    collaboration = "KAGRA",
    title = "{KAGRA: 2.5 Generation Interferometric Gravitational Wave Detector}",
    eprint = "1811.08079",
    archivePrefix = "arXiv",
    primaryClass = "gr-qc",
    reportNumber = "JGW-P1809243",
    doi = "10.1038/s41550-018-0658-y",
    journal = "Nature Astron.",
    volume = "3",
    number = "1",
    pages = "35--40",
    year = "2019"
}

@article{LIGOScientific:2020iuh,
    author = "Abbott, R. and others",
    collaboration = "LIGO Scientific, Virgo",
    title = "{GW190521: A Binary Black Hole Merger with a Total Mass of $150  M_{\odot}$}",
    eprint = "2009.01075",
    archivePrefix = "arXiv",
    primaryClass = "gr-qc",
    doi = "10.1103/PhysRevLett.125.101102",
    journal = "Phys. Rev. Lett.",
    volume = "125",
    number = "10",
    pages = "101102",
    year = "2020"
}

@article{LIGOScientific:2025rsn,
    author = "Abac, A. G. and others",
    collaboration = "LIGO Scientific, VIRGO, KAGRA",
    title = "{GW231123: A Binary Black Hole Merger with Total Mass 190--265 M$_\odot$}",
    eprint = "2507.08219",
    archivePrefix = "arXiv",
    primaryClass = "astro-ph.HE",
    reportNumber = "DCC: P2500026-v6, DCC: P2500026-v8",
    doi = "10.3847/2041-8213/ae0c9c",
    journal = "Astrophys. J. Lett.",
    volume = "993",
    number = "1",
    pages = "L25",
    year = "2025"
}

@article{LIGOScientific:2025brd,
    author = "Abac, A. G. and others",
    collaboration = "LIGO Scientific, Virgo, KAGRA",
    title = "{GW241011 and GW241110: Exploring Binary Formation and Fundamental Physics with Asymmetric, High-spin Black Hole Coalescences}",
    eprint = "2510.26931",
    archivePrefix = "arXiv",
    primaryClass = "astro-ph.HE",
    reportNumber = "LIGO-P2500402",
    doi = "10.3847/2041-8213/ae0d54",
    journal = "Astrophys. J. Lett.",
    volume = "993",
    number = "1",
    pages = "L21",
    year = "2025"
}

@article{Li:2025iux,
    author = "Li, Yin-Jie and Wang, Yuan-Zhu and Tang, Shao-Peng and Fan, Yi-Zhong",
    title = "{Aligned Hierarchical Black Hole Mergers in AGN disks revealed by GWTC-4}",
    eprint = "2509.23897",
    archivePrefix = "arXiv",
    primaryClass = "astro-ph.HE",
    month = "9",
    year = "2025",
    journal = ""
}

@article{Alvarez:2024dpd,
    author = "{\'A}lvarez, Carlos Ara{\'u}jo and Wong, Henry W. Y. and Liu, Anna and Calder{\'o}n Bustillo, Juan",
    title = "{Kicking Time Back in Black Hole Mergers: Ancestral Masses, Spins, Birth Recoils, and Hierarchical-formation Viability of GW190521}",
    eprint = "2404.00720",
    archivePrefix = "arXiv",
    primaryClass = "astro-ph.HE",
    reportNumber = "LIGO-DCC P2400073",
    doi = "10.3847/1538-4357/ad90a9",
    journal = "Astrophys. J.",
    volume = "977",
    number = "2",
    pages = "220",
    year = "2024"
}

@article{Gayathri:2020coq,
    author = "Gayathri, V. and Healy, J. and Lange, J. and O'Brien, B. and Szczepanczyk, M. and Bartos, Imre and Campanelli, M. and Klimenko, S. and Lousto, C. O. and O'Shaughnessy, R.",
    title = "{Eccentricity estimate for black hole mergers with numerical relativity simulations}",
    eprint = "2009.05461",
    archivePrefix = "arXiv",
    primaryClass = "astro-ph.HE",
    doi = "10.1038/s41550-021-01568-w",
    journal = "Nature Astron.",
    volume = "6",
    number = "3",
    pages = "344--349",
    year = "2022"
}

@article{Gamba:2021gap,
    author = "Gamba, Rossella and Breschi, Matteo and Carullo, Gregorio and Albanesi, Simone and Rettegno, Piero and Bernuzzi, Sebastiano and Nagar, Alessandro",
    title = "{GW190521 as a dynamical capture of two nonspinning black holes}",
    eprint = "2106.05575",
    archivePrefix = "arXiv",
    primaryClass = "gr-qc",
    doi = "10.1038/s41550-022-01813-w",
    journal = "Nature Astron.",
    volume = "7",
    number = "1",
    pages = "11--17",
    year = "2023"
}

@article{Romero-Shaw:2022xko,
    author = "Romero-Shaw, Isobel M. and Lasky, Paul D. and Thrane, Eric",
    title = "{Four Eccentric Mergers Increase the Evidence that LIGO\textendash{}Virgo\textendash{}KAGRA\textquoteright{}s Binary Black Holes Form Dynamically}",
    eprint = "2206.14695",
    archivePrefix = "arXiv",
    primaryClass = "astro-ph.HE",
    doi = "10.3847/1538-4357/ac9798",
    journal = "Astrophys. J.",
    volume = "940",
    number = "2",
    pages = "171",
    year = "2022"
}

@article{Gupte:2024jfe,
    author = "Gupte, Nihar and others",
    title = "{Evidence for eccentricity in the population of binary black holes observed by LIGO-Virgo-KAGRA}",
    eprint = "2404.14286",
    archivePrefix = "arXiv",
    primaryClass = "gr-qc",
    doi = "10.1103/vpyp-nvfp",
    journal = "Phys. Rev. D",
    volume = "112",
    number = "10",
    pages = "104045",
    year = "2025"
}

@article{Romero-Shaw:2025vbc,
    author = "Romero-Shaw, Isobel and Stegmann, Jakob and Tagawa, Hiromichi and Gerosa, Davide and Samsing, Johan and Gupte, Nihar and Green, Stephen R.",
    title = "{GW200208{\_}222617 as an eccentric black-hole binary merger: Properties and astrophysical implications}",
    eprint = "2506.17105",
    archivePrefix = "arXiv",
    primaryClass = "astro-ph.HE",
    doi = "10.1103/jj7m-x66y",
    journal = "Phys. Rev. D",
    volume = "112",
    number = "6",
    pages = "063052",
    year = "2025"
}

@article{Planas:2025jny,
    author = "Planas, Maria de Lluc and Ramos-Buades, Antoni and Garc{\'\i}a-Quir{\'o}s, Cecilio and Estell{\'e}s, H{\'e}ctor and Husa, Sascha and Haney, Maria",
    title = "{Reanalysis of binary black hole gravitational wave events for orbital eccentricity signatures}",
    eprint = "2504.15833",
    archivePrefix = "arXiv",
    primaryClass = "gr-qc",
    doi = "10.1103/cv75-y8dr",
    journal = "Phys. Rev. D",
    volume = "112",
    number = "12",
    pages = "123004",
    year = "2025"
}

@article{Planas:2025plq,
    author = "Planas, Maria de Lluc and Husa, Sascha and Ramos-Buades, Antoni and Valencia, Jorge",
    title = "{First Eccentric Inspiral{\textendash}Merger{\textendash}Ringdown Analysis of Neutron Star{\textendash}Black Hole Mergers}",
    eprint = "2506.01760",
    archivePrefix = "arXiv",
    primaryClass = "astro-ph.HE",
    doi = "10.3847/1538-4357/ae1d7d",
    journal = "Astrophys. J.",
    volume = "995",
    number = "1",
    pages = "47",
    year = "2025"
}

@article{Phukon:2025cky,
    author = "Phukon, Khun Sang and Schmidt, Patricia and Morras, Gonzalo and Pratten, Geraint",
    title = "{Detection of GW200105 with a targeted eccentric search}",
    eprint = "2512.10803",
    archivePrefix = "arXiv",
    primaryClass = "gr-qc",
    reportNumber = "LIGO-P2500673",
    doi = "10.1103/cmtb-5q4b",
    journal = "Phys. Rev. D",
    volume = "113",
    number = "10",
    pages = "103023",
    year = "2026"
}

@article{Jan:2025fps,
    author = "Jan, Aasim and Tsao, Bing-Jyun and O'Shaughnessy, Richard and Shoemaker, Deirdre and Laguna, Pablo",
    title = "{GW200105: A detailed study of eccentricity in the neutron star-black hole binary}",
    eprint = "2508.12460",
    archivePrefix = "arXiv",
    primaryClass = "gr-qc",
    doi = "10.1103/zjmc-117s",
    journal = "Phys. Rev. D",
    volume = "113",
    number = "2",
    pages = "024018",
    year = "2026"
}

@article{Kacanja:2025kpr,
    author = "Kacanja, Keisi and Soni, Kanchan and Nitz, Alexander Harvey",
    title = "{Eccentricity signatures in LIGO-Virgo-KAGRA{\textquoteright}s binary neutron star and neutron-star black holes}",
    eprint = "2508.00179",
    archivePrefix = "arXiv",
    primaryClass = "gr-qc",
    doi = "10.1103/jnsc-783p",
    journal = "Phys. Rev. D",
    volume = "112",
    number = "12",
    pages = "122007",
    year = "2025"
}

@article{Tiwari:2025fua,
    author = "Tiwari, Avinash and Bhat, Sajad A. and Shaikh, Md Arif and Kapadia, Shasvath J.",
    title = "{Testing the Nature of GW200105 by Probing the Frequency Evolution of Eccentricity}",
    eprint = "2509.26152",
    archivePrefix = "arXiv",
    primaryClass = "astro-ph.HE",
    doi = "10.3847/1538-4357/ae1d74",
    journal = "Astrophys. J.",
    volume = "995",
    number = "1",
    pages = "48",
    year = "2025"
}

@article{LIGOScientific:2025pvj,
    author = "Abac, A. G. and others",
    collaboration = "LIGO Scientific, VIRGO, KAGRA",
    title = "{GWTC-4.0: Population Properties of Merging Compact Binaries}",
    eprint = "2508.18083",
    archivePrefix = "arXiv",
    primaryClass = "astro-ph.HE",
    reportNumber = "LIGO-P2400004",
    month = "8",
    year = "2025",
    journal =""
}

@article{Romero-Shaw:2025otx,
    author = {Romero-Shaw, Isobel and Stegmann, Jakob and Morras, Gonzalo and Dorozsmai, Andris and Zevin, Michael},
    title = {Astrophysical implications of eccentricity in gravitational waves from neutron star-black hole binaries},
    journal = {Monthly Notices of the Royal Astronomical Society},
    volume = {547},
    number = {2},
    pages = {stag323},
    year = {2026},
    month = {04},
    issn = {0035-8711},
    doi = {10.1093/mnras/stag323},
    url = {https://doi.org/10.1093/mnras/stag323},
    eprint = "2512.16289",
    archivePrefix = "arXiv",
}

@article{Stegmann:2025zkb,
    author = "Stegmann, Jakob and Antonini, Fabio and Olejak, Aleksandra and Biscoveanu, Sylvia and Raymond, Vivien and Rinaldi, Stefano and Flanagan, Elizabeth",
    title = "{Gravitational-wave Observations Suggest Most Black Hole Mergers Form in Triples}",
    eprint = "2512.15873",
    archivePrefix = "arXiv",
    primaryClass = "astro-ph.HE",
    doi = "10.3847/2041-8213/ae52ec",
    journal = "Astrophys. J. Lett.",
    volume = "1000",
    number = "2",
    pages = "L59",
    year = "2026"
}

@article{ET:2019dnz,
    author = "Maggiore, Michele and others",
    collaboration = "ET",
    title = "{Science Case for the Einstein Telescope}",
    eprint = "1912.02622",
    archivePrefix = "arXiv",
    primaryClass = "astro-ph.CO",
    doi = "10.1088/1475-7516/2020/03/050",
    journal = "JCAP",
    volume = "03",
    pages = "050",
    year = "2020"
}

@article{Branchesi:2023mws,
    author = "Branchesi, Marica and others",
    title = "{Science with the Einstein Telescope: a comparison of different designs}",
    eprint = "2303.15923",
    archivePrefix = "arXiv",
    primaryClass = "gr-qc",
    reportNumber = "ET-0084A-23",
    doi = "10.1088/1475-7516/2023/07/068",
    journal = "JCAP",
    volume = "07",
    pages = "068",
    year = "2023"
}

@article{Reitze:2019iox,
    author = "Reitze, David and others",
    title = "{Cosmic Explorer: The U.S. Contribution to Gravitational-Wave Astronomy beyond LIGO}",
    eprint = "1907.04833",
    archivePrefix = "arXiv",
    primaryClass = "astro-ph.IM",
    reportNumber = "LIGO-P1900316",
    journal = "Bull. Am. Astron. Soc.",
    volume = "51",
    number = "7",
    pages = "035",
    year = "2019"
}

@article{Evans:2021gyd,
    author = "Evans, Matthew and others",
    title = "{A Horizon Study for Cosmic Explorer: Science, Observatories, and Community}",
    eprint = "2109.09882",
    archivePrefix = "arXiv",
    primaryClass = "astro-ph.IM",
    reportNumber = "CE-P2100003-v7, Cosmic Explorer technical report CE-P2100003-v6",
    month = "9",
    year = "2021",
    journal = "",
}

@article{LISA:2017pwj,
    author = "Amaro-Seoane, Pau and others",
    collaboration = "LISA",
    title = "{Laser Interferometer Space Antenna}",
    eprint = "1702.00786",
    archivePrefix = "arXiv",
    primaryClass = "astro-ph.IM",
    month = "2",
    year = "2017",
    journal = "",
}

@article{LISA:2024hlh,
    author = "Colpi, Monica and others",
    collaboration = "LISA",
    title = "{LISA Definition Study Report}",
    eprint = "2402.07571",
    archivePrefix = "arXiv",
    primaryClass = "astro-ph.CO",
    month = "2",
    year = "2024",
    journal = "",
}

@article{LIGOScientific:2018mvr,
    author = "Abbott, B. P. and others",
    collaboration = "LIGO Scientific, Virgo",
    title = "{GWTC-1: A Gravitational-Wave Transient Catalog of Compact Binary Mergers Observed by LIGO and Virgo during the First and Second Observing Runs}",
    eprint = "1811.12907",
    archivePrefix = "arXiv",
    primaryClass = "astro-ph.HE",
    reportNumber = "LIGO-P1800307",
    doi = "10.1103/PhysRevX.9.031040",
    journal = "Phys. Rev. X",
    volume = "9",
    number = "3",
    pages = "031040",
    year = "2019"
}

@article{LIGOScientific:2020ibl,
    author = "Abbott, R. and others",
    collaboration = "LIGO Scientific, Virgo",
    title = "{GWTC-2: Compact Binary Coalescences Observed by LIGO and Virgo During the First Half of the Third Observing Run}",
    eprint = "2010.14527",
    archivePrefix = "arXiv",
    primaryClass = "gr-qc",
    reportNumber = "P2000061",
    doi = "10.1103/PhysRevX.11.021053",
    journal = "Phys. Rev. X",
    volume = "11",
    pages = "021053",
    year = "2021"
}

@article{KAGRA:2021vkt,
    author = "Abbott, R. and others",
    collaboration = "KAGRA, VIRGO, LIGO Scientific",
    title = "{GWTC-3: Compact Binary Coalescences Observed by LIGO and Virgo during the Second Part of the Third Observing Run}",
    eprint = "2111.03606",
    archivePrefix = "arXiv",
    primaryClass = "gr-qc",
    reportNumber = "LIGO-P2000318",
    doi = "10.1103/PhysRevX.13.041039",
    journal = "Phys. Rev. X",
    volume = "13",
    number = "4",
    pages = "041039",
    year = "2023"
}

@article{LIGOScientific:2025slb,
    author = "Abac, A. G. and others",
    collaboration = "LIGO Scientific, VIRGO, KAGRA",
    title = "{GWTC-4.0: Updating the Gravitational-Wave Transient Catalog with Observations from the First Part of the Fourth LIGO-Virgo-KAGRA Observing Run}",
    eprint = "2508.18082",
    archivePrefix = "arXiv",
    primaryClass = "gr-qc",
    reportNumber = "LIGO-P2400386",
    month = "8",
    year = "2025",
    journal = "",
}

@article{Lidov:1962,
    author = "Lidov, M. L.",
    title = "The evolution of orbits of artificial satellites of planets under the action of gravitational perturbations of external bodies",
    journal = "Planet. Space Sci.",
    volume = "9",
    pages = "719--759",
    year = "1962",
    doi = "10.1016/0032-0633(62)90054-3"
}

@article{Kozai:1962,
    author = "Kozai, Yoshihide",
    title = "Secular perturbations of asteroids with high inclination and eccentricity",
    journal = "Astron. J.",
    volume = "67",
    pages = "591--598",
    year = "1962",
    doi = "10.1086/108790"
}

@article{Zeipel:1910,
    author = "von Zeipel, H.",
    title = "{Sur l'application des s\'eries de M. Lindstedt \`a l'\'etude du mouvement des com\`etes p\'eriodiques}",
    journal = "Astron. Nachr.",
    volume = "183",
    pages = "34",
    year = "1910",
    doi = "10.1002/asna.19101830302"
}

@article{ET:2025xjr,
    author = "Abac, Adrian and others",
    collaboration = "ET",
    title = "{The Science of the Einstein Telescope}",
    eprint = "2503.12263",
    archivePrefix = "arXiv",
    primaryClass = "gr-qc",
    reportNumber = "ET-0036C-25",
    doi = "10.1088/1475-7516/2026/03/081",
    journal = "JCAP",
    volume = "03",
    pages = "081",
    year = "2026"
}

@article{Morras:2025xfu,
    author = "Morras, Gonzalo and Pratten, Geraint and Schmidt, Patricia",
    title = "{Orbital eccentricity in a neutron star - black hole merger}",
    eprint = "2503.15393",
    archivePrefix = "arXiv",
    primaryClass = "astro-ph.HE",
    reportNumber = "LIGO-DCC P2500105",
    doi = "10.3847/2041-8213/ae474c",
    journal = "Astrophys. J. Lett.",
    volume = "1000",
    number = "1",
    pages = "L2",
    year = "2026"
}

@article{Gerosa:2021mno,
    author = "Gerosa, Davide and Fishbach, Maya",
    title = "{Hierarchical mergers of stellar-mass black holes and their gravitational-wave signatures}",
    eprint = "2105.03439",
    archivePrefix = "arXiv",
    primaryClass = "astro-ph.HE",
    doi = "10.1038/s41550-021-01398-w",
    journal = "Nature Astron.",
    volume = "5",
    number = "8",
    pages = "749--760",
    year = "2021"
}

@article{Stegmann:2025clo,
    author = "Stegmann, Jakob and Klencki, Jakub",
    title = "{Orbital Eccentricity and Spin{\textendash}Orbit Misalignment Are Evidence that Neutron Star{\textendash}Black Hole Mergers Form through Triple Star Evolution}",
    eprint = "2506.09121",
    archivePrefix = "arXiv",
    primaryClass = "astro-ph.HE",
    doi = "10.3847/2041-8213/ae055b",
    journal = "Astrophys. J. Lett.",
    volume = "991",
    number = "2",
    pages = "L54",
    year = "2025"
}

@article{Zi:2023geb,
    author = "Zi, Tieguang and Ye, Chang-Qing and Li, Peng-Cheng",
    title = "{Detecting the tidal heating with the generic extreme mass-ratio inspirals}",
    eprint = "2311.15532",
    archivePrefix = "arXiv",
    primaryClass = "gr-qc",
    doi = "10.1088/1475-7516/2024/10/066",
    journal = "JCAP",
    volume = "10",
    pages = "066",
    year = "2024"
}

@article{Nishimura:2026nse,
    author = "Nishimura, Nami and Buonanno, Alessandra and Faggioli, Guglielmo and van de Meent, Maarten and Khanna, Gaurav",
    title = "{Advancing the Effective-One-Body Framework in the Test-Mass Limit}",
    eprint = "2603.05601",
    archivePrefix = "arXiv",
    primaryClass = "gr-qc",
    month = "3",
    year = "2026",
    journal = ""
}

@article{Taracchini:2013wfa,
    author = "Taracchini, Andrea and Buonanno, Alessandra and Hughes, Scott A. and Khanna, Gaurav",
    title = "{Modeling the horizon-absorbed gravitational flux for equatorial-circular orbits in Kerr spacetime}",
    eprint = "1305.2184",
    archivePrefix = "arXiv",
    primaryClass = "gr-qc",
    doi = "10.1103/PhysRevD.88.044001",
    journal = "Phys. Rev. D",
    volume = "88",
    pages = "044001",
    year = "2013",
    note = "[Erratum: Phys.Rev.D 88, 109903 (2013)]"
}

@article{Barack:2018yvs,
    author = "Barack, Leor and Pound, Adam",
    title = "{Self-force and radiation reaction in general relativity}",
    eprint = "1805.10385",
    archivePrefix = "arXiv",
    primaryClass = "gr-qc",
    doi = "10.1088/1361-6633/aae552",
    journal = "Rept. Prog. Phys.",
    volume = "82",
    number = "1",
    pages = "016904",
    year = "2019"
}

@article{Pound:2021qin,
    author = "Pound, Adam and Wardell, Barry",
    title = "{Black hole perturbation theory and gravitational self-force}",
    eprint = "2101.04592",
    archivePrefix = "arXiv",
    primaryClass = "gr-qc",
    doi = "10.1007/978-981-15-4702-7\_38-1",
    month = "1",
    year = "2021",
    journal = ""
}

@article{Chatziioannou:2016kem,
    author = "Chatziioannou, Katerina and Poisson, Eric and Yunes, Nicolas",
    title = "{Improved next-to-leading order tidal heating and torquing of a Kerr black hole}",
    eprint = "1608.02899",
    archivePrefix = "arXiv",
    primaryClass = "gr-qc",
    doi = "10.1103/PhysRevD.94.084043",
    journal = "Phys. Rev. D",
    volume = "94",
    number = "8",
    pages = "084043",
    year = "2016"
}

@article{Price:1986yy,
    author = "Price, R. H. and Thorne, K. S.",
    title = "{Membrane Viewpoint on Black Holes: Properties and Evolution of the Stretched Horizon}",
    doi = "10.1103/PhysRevD.33.915",
    journal = "Phys. Rev. D",
    volume = "33",
    pages = "915--941",
    year = "1986"
}

@article{Bernuzzi:2012ku,
    author = "Bernuzzi, Sebastiano and Nagar, Alessandro and Zenginoglu, Anil",
    title = "{Horizon-absorption effects in coalescing black-hole binaries: An effective-one-body study of the non-spinning case}",
    eprint = "1207.0769",
    archivePrefix = "arXiv",
    primaryClass = "gr-qc",
    doi = "10.1103/PhysRevD.86.104038",
    journal = "Phys. Rev. D",
    volume = "86",
    pages = "104038",
    year = "2012"
}

@article{Jaraba:2021ces,
    author = "Jaraba, Santiago and Garcia-Bellido, Juan",
    title = "{Black hole induced spins from hyperbolic encounters in dense clusters}",
    eprint = "2106.01436",
    archivePrefix = "arXiv",
    primaryClass = "gr-qc",
    reportNumber = "IFT-UAM/CSIC-21-68",
    doi = "10.1016/j.dark.2021.100882",
    journal = "Phys. Dark Univ.",
    volume = "34",
    pages = "100882",
    year = "2021"
}

@article{Nelson:2019czq,
    author = "Nelson, Patrick E. and Etienne, Zachariah B. and McWilliams, Sean T. and Nguyen, Viviana",
    title = "{Induced Spins from Scattering Experiments of Initially Nonspinning Black Holes}",
    eprint = "1909.08621",
    archivePrefix = "arXiv",
    primaryClass = "gr-qc",
    doi = "10.1103/PhysRevD.100.124045",
    journal = "Phys. Rev. D",
    volume = "100",
    number = "12",
    pages = "124045",
    year = "2019"
}

@article{Rodriguez-Monteverde:2024tnt,
    author = "Rodr{\'\i}guez-Monteverde, Jorge L. and Jaraba, Santiago and Garc{\'\i}a-Bellido, Juan",
    title = "{Spin induction from scattering of two spinning black holes in dense clusters}",
    eprint = "2410.11634",
    archivePrefix = "arXiv",
    primaryClass = "gr-qc",
    reportNumber = "IFT-UAM/CSIC-2024-147",
    doi = "10.1016/j.dark.2024.101776",
    journal = "Phys. Dark Univ.",
    volume = "47",
    pages = "101776",
    year = "2025"
}

@article{Honet:2025gge,
    author = {Honet, Lo{\"\i}c and Pound, Adam and Comp{\`e}re, Geoffrey},
    title = "{Hybrid waveform model for asymmetric spinning binaries: Self-force meets post-Newtonian theory}",
    eprint = "2510.16114",
    archivePrefix = "arXiv",
    primaryClass = "gr-qc",
    doi = "10.1103/rhwy-59y2",
    journal = "Phys. Rev. D",
    volume = "113",
    number = "6",
    pages = "064035",
    year = "2026"
}

@article{Honet:2025lmk,
    author = {Honet, Lo{\"\i}c and Mathews, Josh and Comp{\`e}re, Geoffrey and Pound, Adam and Wardell, Barry and Piovano, Gabriel Andres and van de Meent, Maarten and Warburton, Niels},
    title = "{Spin-aligned inspiral waveforms from self-force and post-Newtonian theory}",
    eprint = "2510.16112",
    archivePrefix = "arXiv",
    primaryClass = "gr-qc",
    month = "10",
    year = "2025",
    journal = ""
}

@misc{pyEFPEHM_repo,
  author = {Morras, Gonzalo and Pratten, Geraint and Schmidt, Patricia and Buonanno, Alessandra},
  title  = {pyEFPEHM Code Repository},
  year   = {2026},
  url    = {https://github.com/gw-models/pyEFPEHM},
}

@misc{pybop_repo,
  author = {Morras, Gonzalo},
  title  = {pybop Code Repository},
  year   = {2026},
  url    = {https://github.com/gmorras/pybop},
}

@article{Ramos-Buades:2023ehm,
    author = "Ramos-Buades, Antoni and Buonanno, Alessandra and Estell{\'e}s, H{\'e}ctor and Khalil, Mohammed and Mihaylov, Deyan P. and Ossokine, Serguei and Pompili, Lorenzo and Shiferaw, Mahlet",
    title = "{Next generation of accurate and efficient multipolar precessing-spin effective-one-body waveforms for binary black holes}",
    eprint = "2303.18046",
    archivePrefix = "arXiv",
    primaryClass = "gr-qc",
    doi = "10.1103/PhysRevD.108.124037",
    journal = "Phys. Rev. D",
    volume = "108",
    number = "12",
    pages = "124037",
    year = "2023"
}

@article{Albanesi:2025txj,
    author = "Albanesi, Simone and Gamba, Rossella and Bernuzzi, Sebastiano and Fontbut{\'e}, Joan and Gonzalez, Alejandra and Nagar, Alessandro",
    title = "{Effective-one-body modeling for generic compact binaries with arbitrary orbits}",
    eprint = "2503.14580",
    archivePrefix = "arXiv",
    primaryClass = "gr-qc",
    doi = "10.1103/3snf-w1x7",
    journal = "Phys. Rev. D",
    volume = "112",
    number = "12",
    pages = "L121503",
    year = "2025"
}

@article{Planas:2025feq,
    author = "Planas, Maria de Lluc and Ramos-Buades, Antoni and Garc{\'\i}a-Quir{\'o}s, Cecilio and Estell{\'e}s, H{\'e}ctor and Husa, Sascha and Haney, Maria",
    title = "{Time-domain phenomenological multipolar waveforms for aligned-spin binary black holes in elliptical orbits}",
    eprint = "2503.13062",
    archivePrefix = "arXiv",
    primaryClass = "gr-qc",
    doi = "10.1103/wz3v-b151",
    journal = "Phys. Rev. D",
    volume = "113",
    number = "2",
    pages = "024006",
    year = "2026"
}

@article{Hamilton:2025xru,
    author = "Hamilton, Eleanor and others",
    title = "{Improved gravitational wave model linking precessing inspirals and numerical-relativity-calibrated merger-ringdown}",
    eprint = "2507.02604",
    archivePrefix = "arXiv",
    primaryClass = "gr-qc",
    doi = "10.1103/kxsf-23rr",
    journal = "Phys. Rev. D",
    volume = "113",
    number = "8",
    pages = "084055",
    year = "2026"
}

@article{Paul:2024ujx,
    author = "Paul, Kaushik and Maurya, Akash and Henry, Quentin and Sharma, Kartikey and Satheesh, Pranav and Divyajyoti and Kumar, Prayush and Mishra, Chandra Kant",
    title = "{Eccentric, spinning, inspiral-merger-ringdown waveform model with higher modes for the detection and characterization of binary black holes}",
    eprint = "2409.13866",
    archivePrefix = "arXiv",
    primaryClass = "gr-qc",
    doi = "10.1103/PhysRevD.111.084074",
    journal = "Phys. Rev. D",
    volume = "111",
    number = "8",
    pages = "084074",
    year = "2025"
}

@article{LIGOScientific:2026wfs,
    collaboration = "LIGO Scientific, VIRGO, KAGRA",
    title = "{GWTC-5.0: Observations from the Second Part of the Fourth LIGO-Virgo-KAGRA Observing Run and Updates to the Gravitational-Wave Transient Catalog}",
    eprint = "2605.27225",
    archivePrefix = "arXiv",
    primaryClass = "gr-qc",
    reportNumber = "LIGO-P2600152",
    month = "5",
    year = "2026",
    journal = ""
}

@article{Gamboa:2026jht,
    author = "Gamboa, Aldo and others",
    title = "{Accurate waveforms for generic planar-orbit binary black holes: The multipolar effective-one-body model SEOBNRv6EHM}",
    eprint = "2605.28715",
    archivePrefix = "arXiv",
    primaryClass = "gr-qc",
    month = "5",
    year = "2026",
    journal = ""
}

@article{Jones1998,
  author    = {Donald R. Jones and Matthias Schonlau and William J. Welch},
  title     = {Efficient Global Optimization of Expensive Black-Box Functions},
  journal   = {Journal of Global Optimization},
  year      = {1998},
  volume    = {13},
  number    = {4},
  pages     = {455--492},
  doi       = {10.1023/A:1008306431147},
  url       = {https://doi.org/10.1023/A:1008306431147},
  issn      = {1573-2916},
}

@article{Pompili:2026yxq,
    author = "Pompili, Lorenzo and Gamboa, Aldo and Buonanno, Alessandra",
    title = "{Eccentric and unbound compact binaries in the LIGO-Virgo-KAGRA catalog: parameter estimation and waveform systematics with SEOBNRv6EHM}",
    eprint = "2605.28716",
    archivePrefix = "arXiv",
    primaryClass = "gr-qc",
    month = "5",
    year = "2026",
    journal = ""
}

@article{Lange:2026eqx,
    author = "Lange, Jacob and Chiaramello, Danilo and Lott, Peter and Henshaw, Chad and Nagar, Alessandro and O'Shaughnessy, Richard and Cadonati, Laura",
    title = "{Gravitational Wave Hyperbolic Catalog: Reanalyzing High-Mass Gravitational Wave Signals Using Hyperbolic Waveforms}",
    eprint = "2605.21640",
    archivePrefix = "arXiv",
    primaryClass = "gr-qc",
    month = "5",
    year = "2026",
    journal = ""
}

@article{Comeau:2009bz,
    author = "Comeau, Simon and Poisson, Eric",
    title = "{Tidal interaction of a small black hole in the field of a large Kerr black hole}",
    eprint = "0908.4518",
    archivePrefix = "arXiv",
    primaryClass = "gr-qc",
    doi = "10.1103/PhysRevD.80.087501",
    journal = "Phys. Rev. D",
    volume = "80",
    pages = "087501",
    year = "2009"
}

@article{Poisson:2009qj,
    author = "Poisson, Eric and Vlasov, Igor",
    title = "{Geometry and dynamics of a tidally deformed black hole}",
    eprint = "0910.4311",
    archivePrefix = "arXiv",
    primaryClass = "gr-qc",
    doi = "10.1103/PhysRevD.81.024029",
    journal = "Phys. Rev. D",
    volume = "81",
    pages = "024029",
    year = "2010"
}

@article{Poisson:2014gka,
    author = "Poisson, Eric",
    title = "{Tidal deformation of a slowly rotating black hole}",
    eprint = "1411.4711",
    archivePrefix = "arXiv",
    primaryClass = "gr-qc",
    doi = "10.1103/PhysRevD.91.044004",
    journal = "Phys. Rev. D",
    volume = "91",
    number = "4",
    pages = "044004",
    year = "2015"
}

@article{Poisson:2018qqd,
    author = "Poisson, Eric and Corrigan, Eamonn",
    title = "{Nonrotating black hole in a post-Newtonian tidal environment II}",
    eprint = "1804.01848",
    archivePrefix = "arXiv",
    primaryClass = "gr-qc",
    doi = "10.1103/PhysRevD.97.124048",
    journal = "Phys. Rev. D",
    volume = "97",
    number = "12",
    pages = "124048",
    year = "2018"
}

@article{Balivada:2026bdb,
    author = "Balivada, Anand and Hegade K. R., Abhishek and Yunes, Nicol{\'a}s",
    title = "{A New Spin on Dissipative Tides: First-Post-Newtonian Effects in Compact Binary Inspirals}",
    eprint = "2604.21990",
    archivePrefix = "arXiv",
    primaryClass = "gr-qc",
    month = "4",
    year = "2026",
    journal = ""
}

@article{Poisson:1994yf,
    author = "Poisson, Eric and Sasaki, Misao",
    title = "{Gravitational radiation from a particle in circular orbit around a black hole. 5: Black hole absorption and tail corrections}",
    eprint = "gr-qc/9412027",
    archivePrefix = "arXiv",
    doi = "10.1103/PhysRevD.51.5753",
    journal = "Phys. Rev. D",
    volume = "51",
    pages = "5753--5767",
    year = "1995"
}

@article{Tagoshi:1997jy,
    author = "Tagoshi, Hideyuki and Mano, Shuhei and Takasugi, Eiichi",
    title = "{PostNewtonian expansion of gravitational waves from a particle in circular orbits around a rotating black hole: Effects of black hole absorption}",
    eprint = "gr-qc/9711072",
    archivePrefix = "arXiv",
    reportNumber = "NAOJ-TH-AP-1997-4",
    doi = "10.1143/PTP.98.829",
    journal = "Prog. Theor. Phys.",
    volume = "98",
    pages = "829--850",
    year = "1997"
}

@article{Shah:2014tka,
    author = "Shah, Abhay G.",
    title = "{Gravitational-wave flux for a particle orbiting a Kerr black hole to 20th post-Newtonian order: a numerical approach}",
    eprint = "1403.2697",
    archivePrefix = "arXiv",
    primaryClass = "gr-qc",
    doi = "10.1103/PhysRevD.90.044025",
    journal = "Phys. Rev. D",
    volume = "90",
    number = "4",
    pages = "044025",
    year = "2014"
}

@article{Munna:2023vds,
    author = "Munna, Christopher and Evans, Charles R. and Forseth, Erik",
    title = "{Tidal heating and torquing of the primary black hole in eccentric-orbit, nonspinning, extreme-mass-ratio inspirals to 22PN order}",
    eprint = "2306.12481",
    archivePrefix = "arXiv",
    primaryClass = "gr-qc",
    doi = "10.1103/PhysRevD.108.044039",
    journal = "Phys. Rev. D",
    volume = "108",
    number = "4",
    pages = "044039",
    year = "2023"
}

@article{Press:1972zz,
    author = "Press, William H. and Teukolsky, Saul A.",
    title = "{Floating Orbits, Superradiant Scattering and the Black-hole Bomb}",
    doi = "10.1038/238211a0",
    journal = "Nature",
    volume = "238",
    pages = "211--212",
    year = "1972"
}

@article{Teukolsky:1974yv,
    author = "Teukolsky, S. A. and Press, W. H.",
    title = "{Perturbations of a rotating black hole. III - Interaction of the hole with gravitational and electromagnetic radiation}",
    doi = "10.1086/153180",
    journal = "Astrophys. J.",
    volume = "193",
    pages = "443--461",
    year = "1974"
}

@article{Fujita:2014eta,
    author = "Fujita, Ryuichi",
    title = "{Gravitational Waves from a Particle in Circular Orbits around a Rotating Black Hole to the 11th Post-Newtonian Order}",
    eprint = "1412.5689",
    archivePrefix = "arXiv",
    primaryClass = "gr-qc",
    doi = "10.1093/ptep/ptv012",
    journal = "PTEP",
    volume = "2015",
    number = "3",
    pages = "033E01",
    year = "2015"
}

@article{Jones:2023ugm,
    author = "Jones, Callum R. T. and Ruf, Michael S.",
    title = "{Absorptive effects and classical black hole scattering}",
    eprint = "2310.00069",
    archivePrefix = "arXiv",
    primaryClass = "hep-th",
    doi = "10.1007/JHEP03(2024)015",
    journal = "JHEP",
    volume = "03",
    pages = "015",
    year = "2024"
}

@article{Cipriani:2026myb,
    author = "Cipriani, Andrea and Fucito, Francesco and Heissenberg, Carlo and Morales, Jose Francisco and Russo, Rodolfo",
    title = "{''Waveforms'' at the Horizon}",
    eprint = "2602.05766",
    archivePrefix = "arXiv",
    primaryClass = "gr-qc",
    month = "2",
    year = "2026",
    journal = ""
}

@article{Warburton:2025ymy,
    author = "Warburton, Niels",
    title = "{Gravitational radiation from hyperbolic orbits: Comparison between self-force, post-Minkowskian, post-Newtonian, and numerical relativity results}",
    eprint = "2512.02274",
    archivePrefix = "arXiv",
    primaryClass = "gr-qc",
    doi = "10.1103/v6rz-4f6k",
    journal = "Phys. Rev. D",
    volume = "113",
    number = "8",
    pages = "084059",
    year = "2026"
}

@article{Zhu:2026mhn,
    author = "Zhu, Hengrui and Pretorius, Frans and Stone, James M.",
    title = "{Trapping, Irregular Waveforms, and Efficient Radiation in Ultra-relativistic Black Hole Encounters}",
    eprint = "2604.26253",
    archivePrefix = "arXiv",
    primaryClass = "gr-qc",
    month = "4",
    year = "2026",
    journal = ""
}

@article{Kogan:2025vml,
    author = "Kogan, Healey and Pardoe, Frederick C. L. and Witek, Helvi",
    title = "{Spin-up and mass-gain in hyperbolic encounters of spinning black holes}",
    eprint = "2511.00307",
    archivePrefix = "arXiv",
    primaryClass = "gr-qc",
    doi = "10.1103/j7ch-t96d",
    journal = "Phys. Rev. D",
    volume = "113",
    number = "6",
    pages = "064016",
    year = "2026"
}

@article{Rodriguez-Monteverde:2025rfh,
    author = "Rodr{\'\i}guez-Monteverde, Jorge L. and Jaraba, Santiago and Garc{\'\i}a-Bellido, Juan",
    title = "{Effects of dynamical capture on two equal-mass nonspinning black holes}",
    eprint = "2510.19699",
    archivePrefix = "arXiv",
    primaryClass = "gr-qc",
    reportNumber = "IFT-UAM/CSIC-25-116",
    doi = "10.1103/x977-swv6",
    journal = "Phys. Rev. D",
    volume = "113",
    number = "2",
    pages = "024052",
    year = "2026"
}

@article{Albertini:2022dmc,
    author = "Albertini, Angelica and Nagar, Alessandro and Pound, Adam and Warburton, Niels and Wardell, Barry and Durkan, Leanne and Miller, Jeremy",
    title = "{Comparing second-order gravitational self-force and effective one body waveforms from inspiralling, quasicircular and nonspinning black hole binaries. II. The large-mass-ratio case}",
    eprint = "2208.02055",
    archivePrefix = "arXiv",
    primaryClass = "gr-qc",
    doi = "10.1103/PhysRevD.106.084062",
    journal = "Phys. Rev. D",
    volume = "106",
    number = "8",
    pages = "084062",
    year = "2022"
}

@article{Mukherjee:2025wxa,
    author = "Mukherjee, Samanwaya and Datta, Sayak and Bose, Sukanta and Phukon, Khun Sang",
    title = "{Tidal heating effects in binary black hole mergers}",
    eprint = "2506.22363",
    archivePrefix = "arXiv",
    primaryClass = "gr-qc",
    doi = "10.1103/wydq-hzvb",
    journal = "Phys. Rev. D",
    volume = "113",
    number = "10",
    pages = "104011",
    year = "2026"
}

@article{Datta:2020gem,
    author = "Datta, Sayak and Phukon, Khun Sang and Bose, Sukanta",
    title = "{Recognizing black holes in gravitational-wave observations: Challenges in telling apart impostors in mass-gap binaries}",
    eprint = "2004.05974",
    archivePrefix = "arXiv",
    primaryClass = "gr-qc",
    reportNumber = "LIGO-P2000115",
    doi = "10.1103/PhysRevD.104.084006",
    journal = "Phys. Rev. D",
    volume = "104",
    number = "8",
    pages = "084006",
    year = "2021"
}

@article{Chapman-Bird:2025xtd,
    author = "Chapman-Bird, Christian E. A. and others",
    title = "{Efficient waveforms for asymmetric-mass eccentric equatorial inspirals into rapidly spinning black holes}",
    eprint = "2506.09470",
    archivePrefix = "arXiv",
    primaryClass = "gr-qc",
    doi = "10.1103/scbp-75pf",
    journal = "Phys. Rev. D",
    volume = "112",
    number = "10",
    pages = "104023",
    year = "2025"
}

@article{Nagar:2011aa,
    author = "Nagar, Alessandro and Akcay, Sarp",
    title = "{Horizon-absorbed energy flux in circularized, nonspinning black-hole binaries and its effective-one-body representation}",
    eprint = "1112.2840",
    archivePrefix = "arXiv",
    primaryClass = "gr-qc",
    doi = "10.1103/PhysRevD.85.044025",
    journal = "Phys. Rev. D",
    volume = "85",
    pages = "044025",
    year = "2012"
}

@article{Damour:2012ky,
    author = "Damour, Thibault and Nagar, Alessandro and Bernuzzi, Sebastiano",
    title = "{Improved effective-one-body description of coalescing nonspinning black-hole binaries and its numerical-relativity completion}",
    eprint = "1212.4357",
    archivePrefix = "arXiv",
    primaryClass = "gr-qc",
    doi = "10.1103/PhysRevD.87.084035",
    journal = "Phys. Rev. D",
    volume = "87",
    number = "8",
    pages = "084035",
    year = "2013"
}

@article{Pratten:2020fqn,
    author = "Pratten, Geraint and Husa, Sascha and Garcia-Quiros, Cecilio and Colleoni, Marta and Ramos-Buades, Antoni and Estelles, Hector and Jaume, Rafel",
    title = "{Setting the cornerstone for a family of models for gravitational waves from compact binaries: The dominant harmonic for nonprecessing quasicircular black holes}",
    eprint = "2001.11412",
    archivePrefix = "arXiv",
    primaryClass = "gr-qc",
    reportNumber = "LIGO-P2000018",
    doi = "10.1103/PhysRevD.102.064001",
    journal = "Phys. Rev. D",
    volume = "102",
    number = "6",
    pages = "064001",
    year = "2020"
}

@article{Estelles:2020twz,
    author = "Estell{\'e}s, H{\'e}ctor and Husa, Sascha and Colleoni, Marta and Keitel, David and Mateu-Lucena, Maite and Garc{\'\i}a-Quir{\'o}s, Cecilio and Ramos-Buades, Antoni and Borchers, Angela",
    title = "{Time-domain phenomenological model of gravitational-wave subdominant harmonics for quasicircular nonprecessing binary black hole coalescences}",
    eprint = "2012.11923",
    archivePrefix = "arXiv",
    primaryClass = "gr-qc",
    doi = "10.1103/PhysRevD.105.084039",
    journal = "Phys. Rev. D",
    volume = "105",
    number = "8",
    pages = "084039",
    year = "2022"
}

@article{Goldberger:2020fot,
    author = "Goldberger, Walter D. and Li, Jingping and Rothstein, Ira Z.",
    title = "{Non-conservative effects on spinning black holes from world-line effective field theory}",
    eprint = "2012.14869",
    archivePrefix = "arXiv",
    primaryClass = "hep-th",
    doi = "10.1007/JHEP06(2021)053",
    journal = "JHEP",
    volume = "06",
    pages = "053",
    year = "2021"
}

\end{document}